\documentclass[format=acmsmall, review=false, screen=true]{acmart}
\usepackage{booktabs} % For formal tables

\usepackage[ruled]{algorithm2e} % For algorithms
\usepackage{subfig}
\usepackage{multirow}

\SetAlFnt{\small}
\SetAlCapFnt{\small}
\SetAlCapNameFnt{\small}
\SetAlCapHSkip{0pt}
\IncMargin{-\parindent}
\acmDOI{0000001.0000001}

\begin{document}
% Title portion
\title{Optimizing the SSD Burst Buffer by Traffic Detection}
 %\titlenote{This is a titlenote}
 %\subtitle{This is a subtitle}
 %\subtitlenote{Subtitle note}

\author{Xuanhua Shi}
%\orcid{1234-5678-9012-3456}
\affiliation{%
  \institution{Huazhong University of Science and Technology}
  \streetaddress{1037 LuoYu East Road}
  \city{Wuhan}
  \state{Hubei}
  \postcode{430074}
  \country{China}}
\email{xhshi@hust.edu.cn}

\author{Wei Liu}
\affiliation{%
  \institution{Huazhong University of Science and Technology}
  \streetaddress{1037 LuoYu East Road}
  \city{Wuhan}
  \state{Hubei}
  \postcode{430074}
  \country{China}}
\email{cccloude@hust.edu.cn}

\author{Ligang He}
\affiliation{%
  \institution{University of Warwick}
  \department{Department of Computer Science}
  %\streetaddress{30 Shuangqing Rd}
  \city{Coventry}
  %\state{Beijing Shi}
  \country{UK}}
\email{ligang.he@warwick.ac.uk}

\author{Hai Jin}
\affiliation{%
  \institution{Huazhong University of Science and Technology}
  \streetaddress{1037 LuoYu East Road}
  \city{Wuhan}
  \state{Hubei}
  \postcode{430074}
  \country{China}}
\email{hjin@hust.edu.cn}

\author{Ming Li}
\affiliation{%
  \institution{Huazhong University of Science and Technology}
  \streetaddress{1037 LuoYu East Road}
  \city{Wuhan}
  \state{Hubei}
  \postcode{430074}
  \country{China}}
\email{limingcs@hust.edu.cn}

\author{Yong Chen}
\affiliation{%
  \institution{Texas Tech University}
  \department{Department of Computer Science}
  \city{Lubbock}
  \state{Texas}
  \postcode{22903}
  \country{USA}}
\email{yong.chen@ttu.edu}

\begin{abstract}
Currently, HPC storage systems still use \textit{hard disk drive} (HDD) as their dominant storage device. \textit{Solid state drive} (SSD) is widely deployed as the buffer to HDDs. \textit{Burst buffer} has also been proposed to manage the SSD buffering of bursty write requests. Although burst buffer can improve I/O performance in many cases, we find that it has some limitations such as requiring large SSD capacity and harmonious overlapping between computation phase and data flushing phase.

In this paper, we propose a scheme, called SSDUP+\footnote{The previous version of this scheme, called SSDUP, has been published in ICS2017~\cite{xuanhua2017ssdup}. The extension of SSDUP+ from SSDUP is summarized in subsection \ref{extension} when we discuss the related work.}. SSDUP+ aims to improve the burst buffer by addressing the above limitations.
First, in order to reduce the demand for the SSD capacity, we develop a novel method
to detect and quantify the data randomness in the write traffic. Further, an adaptive
algorithm is proposed to classify the random writes dynamically. By doing so, much less SSD capacity is required to achieve the similar performance as other burst buffer schemes. 
%Based on this method, only the random writes are buffered to SSD, while other writes are deemed sequential and propagated to HDDs directly. 
Next, in order to overcome the difficulty of perfectly overlapping the computation phase and the flushing phase, we propose a pipeline mechanism for the SSD buffer, in which data buffering and flushing are performed in pipeline. In addition, in order to improve the I/O throughput, we adopt a traffic-aware flushing strategy to reduce the I/O interference in HDD. 
Finally, in order to further improve the performance of buffering random writes in SSD, SSDUP+ transforms the random writes to sequential writes in SSD by storing the data with a log structure. Further, SSDUP+ uses the AVL tree structure to store the sequence information of the data. 

We have implemented a prototype of SSDUP+ based on OrangeFS and conducted extensive experiments. The experimental results show that our proposed SSDUP+ can save an average of 50\% SSD space, while delivering almost the same performance as other common burst buffer schemes. In addition, SSDUP+ can save about 20\% SSD space compared with the previous version of this work, SSDUP, while achieving 20\%-30\% higher I/O throughput than SSDUP.
\end{abstract}

%\begin{CCSXML}
%<ccs2012>
%<concept>
%<concept_id>10002951.10003152.10003517</concept_id>
%<concept_desc>Information systems~Information storage systems</concept_desc>
%<concept_significance>500</concept_significance>
%</concept>
%<concept>
%<concept_id>10002951.10003152.10003517</concept_id>
%<concept_desc>Information systems~Storage Architecture</concept_desc>
%<concept_significance>500</concept_significance>
%</concept>
%</ccs2012>
%\end{CCSXML}

%\ccsdesc[500]{Information systems~Information storage systems}
%\ccsdesc[300]{Information systems~Storage Architecture}



\maketitle

 \section{Introduction}

While \textit{high-performance computing} (HPC) systems are moving towards the exascale era,
the I/O performance remains one of the main bottlenecks, especially for many
data-intensive scientific applications. With the continuous and rapid growth of data volume, the storage demand of many HPC systems has reached the petabyte level~\cite{luu2015multiplatform}. Under this level of storage demand, \textit{hard disk drives} (HDDs) are still
used as the main permanent storage devices in HPC systems, partly because of their
low cost and partly because HDDs can offer high bandwidth when accessing large continuous data chunks. However, HDDs have a major drawback: they suffer from the poor performance when the data are accessed randomly because of the slow mechanical movement of disk heads.

New storage devices such as \textit{solid state drives} (SSDs) have been widely deployed in the HPC environment because of their near-zero seek latency and superior performance (especially for random accesses)~\cite{SSDCOVER}~\cite{chen2011hystor}~\cite{he2014s4d}~\cite{zhang2013ibridge}~\cite{huang2013liu}. However, SSDs are much more expensive than HDDs. Therefore, it is not a cost effective solution to use SSDs as the sole storage devices in large-scale production HPC systems, not to mention the technical limitations of SSDs, such as the issues of wear-out and limited lifetime. A popular solution to address the problem of random data access to HDDs is using the SSDs to buffer the data streams between HDDs and computing nodes, which has been adopted by many production supercomputers (e.g., Sunway TaihuLight and Tianhe-2).

Another feature of HPC systems is that most applications running on HPC systems are write-heavy. The exemplar applications include those from the domains of climate science, physics, earth science etc., which mainly perform the numerical simulations. Moreover, the write requests issued by these applications are bursty because the applications often generate and store a large amount of intermediate results~\cite{liu2014automatic}~\cite{liu2012role}. Further, in order to provide failure protections the concurrent processes of an application often perform checkpointing simultaneously and dump the in-memory data to the permanent storage, which generates the bursty write operations as well. The bursty random writes to HDDs could significantly degrade the performance of data-intensive applications running on HPC systems.
 
In order to address the above issue, \textit{Burst Buffer} (BB)~\cite{liu2012role} has been introduced, which uses an SSD buffer as an intermediate layer between the compute nodes and the HDD-based storage servers to absorb the bursty write
requests. In BB, the data generated by an application are first written to the SSD buffer, which effectively absorbs the bursty writes of the application since SSD offers very low latency. While the data in the SSD buffer are flushed to HDDs, the HPC system continues to process next computation phase. Although BB significantly improves the write performance of HPC applications in many cases, we find that the workings of BB rely on the following two conditions to achieve the expected performance improvement, both of which cannot be easily met.

\textbf{Large SSD capacity}. BB requires the SSD buffer to have the adequate capacity to store all data generated by the computation phase of an application. Otherwise, when the SSD buffer is fully occupied by the write data, the application has two options depending on the implementations of BB: writing the future data to HDD directly or blocking the incoming I/O requests until BB becomes available again.  In both options, the overall I/O performance will be dominated by HDD, which violates the purpose of using burst buffer.
%the application has two options depending on the implementations of BB. One is that the application writes the data to HDD directly, which obviously leads to low performance. In the other option, the application waits until the flushing stage is completed (i.e., the existing data in SSD is flushed to HDD) so as to clear the space to accommodate more data. This option causes low performance too because of the feature of SSD writing (erasing the data before writing).
This condition means that the capacity of the SSD buffer has to be no less than the maximum size of the data generated by a computation phase of an application. However, an HPC application can access hundreds of terabytes or even tens of petabytes of data. For instance, the total data transmission of the Earth1 application in the Mira supercomputer has reached about 10PB~\cite{luu2015multiplatform}. On the other hand, although the data size of individual applications has reached the petabyte level, such applications account for only 10\% of all applications running in a typical HPC platform. The data size of the remaining 90\% of the applications are still around several
GBs~\cite{luu2015multiplatform}. There is now a dilemma: shall we use large (therefore expensive) SSDs to meet the need of a small fraction of applications or use moderate-sized (therefore cost effective) SSDs to satisfy most applications but sacrifice the high-demanding applications?

\textbf{Overlapping computation with data flushing}. BB needs to overlap the computing phase with the flushing stage of data writing. It is not easy to meet this condition either. On one hand, it is difficult to predict the duration of a computation phase, which varies greatly subject to many factors. On the other hand, the HPC platforms use job scheduling tools, such as PBS~\cite{PBS}, SGE~\cite{SGE}, and slurm~\cite{slurm}, to manage job submissions and executions.
To the best of our knowledge, these job schedulers do not take
the usage of storage resources into consideration. Therefore, it is very likely that before the previous flushing stage is completed, the computing phase from either the same application or a different application start to write new data. Once the SSD buffer becomes full, the applications once again face the aforementioned two options, both of which unfortunately deliver low performance.

This work strives to address the above difficult issues in BB and develop a traffic-aware SSD burst buffer scheme, called the \textit{SSDUP+}~\footnote{The source code is available at https://github.com/CGCL-codes/SSDUPplus}, for large-scale HPC systems. The following strategies are developed in SSDUP+ to reduce the SSD capacity required to satisfy the performance of bursty, large-scale I/O accesses (or from another perspective, to improve the I/O performance given the same SSD capacity).

First, in order to reduce the demand for the SSD buffer capacity, we select only a proportion of data to write to SSD, while the remaining data are written to HDDs directly without sacrificing the I/O performance. A novel I/O-traffic detection method is developed to detect the data access patterns of the running processes. Only the random access writes are directed to SSD while the remaining writes are deemed as sequential accesses and propagated to HDDs directly. Our I/O-traffic detection method is novel because the current method of detecting the data access pattern from multiple processes is mainly through calling the collective MPI-IO operations in the applications. This \textit{client-side} method has the following limitations. i) These multiple processes are in the scope of the same application. Therefore, it can only detect the data access pattern within a single application, not from different applications. ii) The method can only detect the data access pattern related to the I/O operation that is being invoked currently, not across different I/O invocations. Our I/O-traffic detection component is located at the side of the storage server. In our \textit{server-side} detection method, a novel metric, termed \textit{random factor}, is proposed to detect the randomness of the write streams, no matter they are across different I/O invocations by the same application or across different applications. Furthermore, an adaptive algorithm is proposed to dynamically adjust the threshold of random factor according to the feature of the workload. The threshold is used to determine whether the current requests are buffered in SSD or written to HDD directly. 

Second, in order to overcome the difficulty of accurately predicting the duration of the computing phase and further improve I/O performance, we develop a pipeline mechanism for the SSD buffer. In the pipeline mechanism, the SSD buffer is divided into two halves. While one half is receiving the writing data, the other fully occupied half flushes the data from SSD to HDD. With the pipeline mechanism, a portion of the SSD buffer will have been flushed and therefore ready to accommodate new data even if the amount of data produced by the computation phase is larger than the capacity of SSD buffer. Therefore, when the new computation phase produces new data (either due to the inaccurate prediction of computation phases or I/O unawareness of job schedulers), which will block the application or be written to HDDs directly in the existing implementations of BB, the new data can still be written to the SSD buffer and consequently the I/O performance can be significantly improved. Moreover, when multiple applications access a storage node concurrently, we find that when the data from one application is being flushed from SSD to HDD, other applications may be writing their data to HDD at the same time, which will cause intense I/O interference in HDD and consequently lead to performance degradation. To address this issue, we propose a traffic-aware flushing strategy in SSDUP+. The pipeline module of SSDUP+ dynamically analyzes the current workload and decides a good timing for performing the flushing operation according to the current workload in HDD.

Finally, we only buffer the random writes to SSD. However, SSD is desired for sequential writes due to the features of SSD writing. In order to improve the performance of buffering random writes in SSD. We convert the random writes to sequential writes by storing the writes in a log-structured manner, i.e., appending the data to the end of the buffered files. A drawback of such a log structure is that the original sequence of file requests is lost. In order to address this issue, we propose to use the AVL tree structure to store the information of the file sequences. Our analysis shows that we only need a tiny fraction of extra storage space to store the AVL tree structure. Therefore, we significantly improve the performance of write buffering at a very small storage expense.

In summary, the contributions of this work are as follows:
% itemize
\begin{itemize}
  \item We carefully analyze the common access patterns in HPC systems and propose a method to detect the randomness of data writing, using a new metric called \textit{random factor}. In addition, an adaptive algorithm is proposed to dynamically adjust the threshold of \textit{random factor}, which is used to determine whether the requests are buffered the SSD. Based on the identified access pattern, the I/O traffic is reshaped and random accesses are directed to the SSD to achieve the highly efficient write buffering.
  %which is aware of the I/O bottleneck of the hard disk from a global point of view. We then
  %reshape the I/O traffic based on our algorithm.
  \item We design a pipeline mechanism for the SSD buffer. It handles the buffering stage and the flushing stage in pipeline and can effectively mitigate the negative impact due to the difficulty of precisely overlapping the computation phase and the flushing stage. Moreover, we design a traffic-aware flushing strategy, aiming to avoid the I/O interference between the SSD flushing and the data writing performed by other applications in HDD. 
  \item A log structure is used to transform the random writes to sequential writes in the SSD buffer and an AVL tree structure is further used to maintain the sequence information of the file requests.
  \item We conduct the effectiveness analysis for the proposed schemes. Moreover, we have implemented a prototype of SSDUP+ based on the OrangeFS and carried out the extensive experimental evaluation. The experimental results show that SSDUP+ outperforms existing burst buffering schemes, including the previous version of this work, SSDUP. 
\end{itemize}

The rest of the paper is organized as follows. Section 2 describes the
detailed design of SSDUP+. Evaluations and analyses are presented in Section
3. The related work in recent years and their relation are discuss in Section 4. Finally,  this paper is concluded in Section 5.

% Head 1
\section{SSDUP+}
In this section, we describe the design of SSDUP+ (a traffic-aware SSD burst buffer).
SSDUP+ contains four main modules, as shown in Figure~\ref{figure-architecture},
including a \textit{random access detector} component that identifies random/irregular
I/O accesses, a \textit{data redirector} component that redirects I/O requests to different
devices, a \textit{pipeline} component that orders and schedules I/O requests in pipeline efficiently, and an \textit{AVL tree management} that manages
the buffered data and maintains the data sequence to order the data.
In this section, we first overview SSDUP+, and then present the details
of each component in SSDUP+, including the designed algorithms and the data structures.
% Head 2
\subsection{Overview of SSDUP+}
\begin{figure}[htb]
  \vspace{-0.4cm}
  \centering
  \includegraphics[width=0.38\textwidth, height=5.4cm]{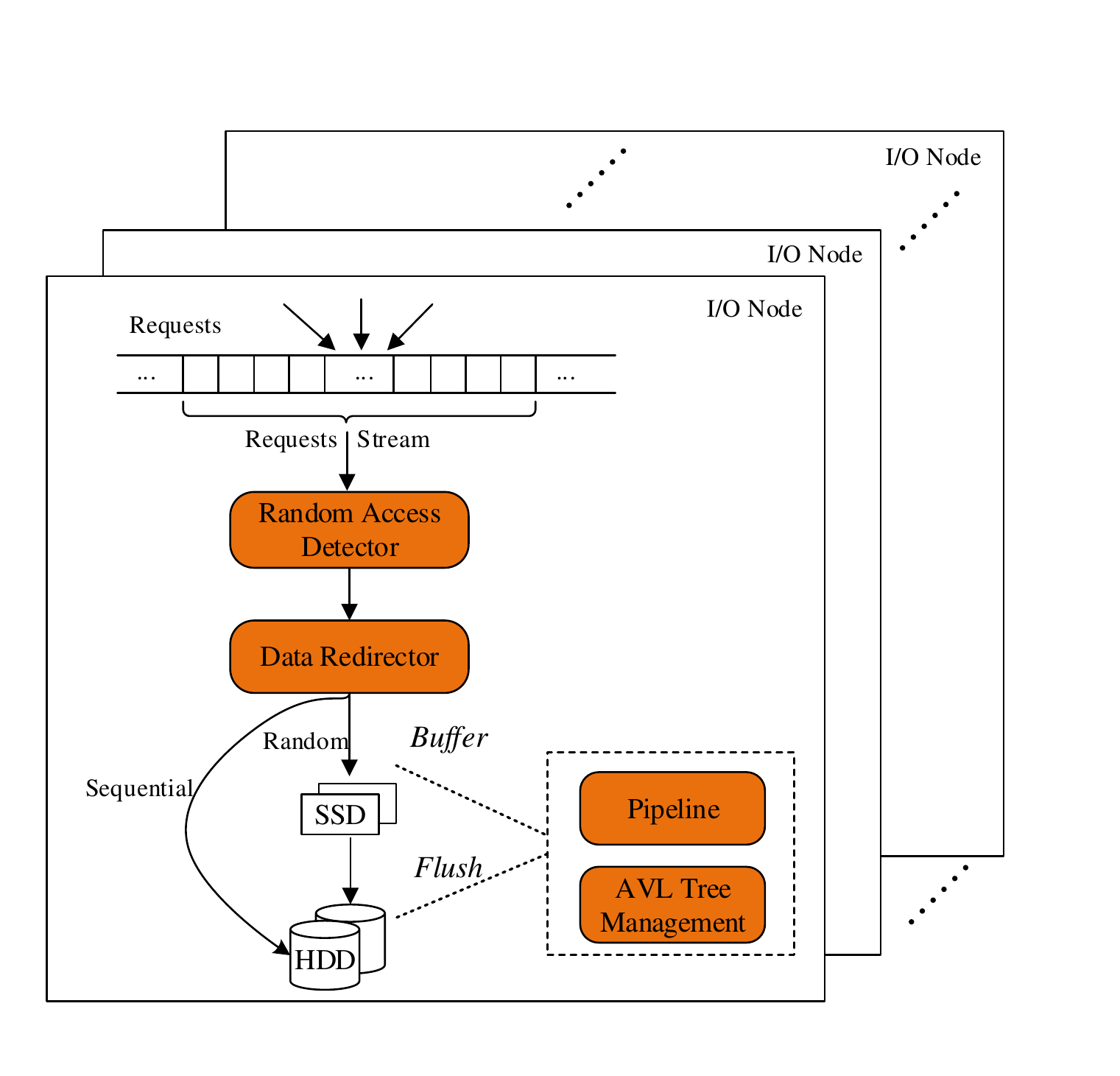}
  \vspace{-0.3cm}
  \caption{The Architecture of SSDUP+}
  \label{figure-architecture}
  \vspace{-0.2cm}
\end{figure}

Although SSDUP+ is implemented as a part of OrangeFS~\cite{orangefs} (the latest version of PVFS2) in this work, the methodology used to design SSDUP+ is generic. The algorithms and the data structures designed in SSDUP+ can also be applied to other file systems.

OrangeFS adopts a client/server model. It has been widely used not only as
an experimental platform, but also a production platform in the HPC area.
In OrangeFS, files are striped across multiple I/O nodes for concurrent accesses. When issuing an I/O request,
the client first communicates with the metadata server to retrieve the data location,
and then issues multiple sub-requests to I/O nodes where the data are located.

Figure~\ref{figure-architecture} shows the architecture of SSDUP+. SSDUP+ resides in each I/O node
and integrates into the \textit{pvfs2-server} daemon. SSDUP+ in different I/O nodes does not need to communicate with
each other. When the requests arrive at an I/O node, SSDUP+ groups the requests into blocks. Each block contains a sequence of requests, which is also called a \textit{request stream}. The block size (e.g., 128 bytes) is a system parameter.
Next, SSDUP+ reshapes the request stream with the help of four components. The \textit{random access detector} is responsible for determining whether the subsequent request stream should be written to HDD or SSD. The \textit{data redirector} is responsible for sending the data to the dedicated devices based on the result of the random access detector. The \textit{pipeline} component handles the data buffering stage and the flushing stage, and maintains
the sufficient SSD space for the incoming requests. The \textit{AVL tree management} component
manages the metadata of the buffered data and sorts the random data for better flushing performance.

There exist different architectures for implementing the burst buffer. Two typical architectures are I/O server-oriented architecture~\cite{wang2017metakv}~\cite{li2013zht} and computing node-oriented architecture~\cite{wang2016burstfs}~\cite{zhao2014fusionfs}. In the I/O server-oriented architecture, the burst buffer system is deployed as an I/O server, which accepts the I/O requests from clients (e.g., computing nodes), while in the computing node-oriented architecture, the burst buffer mechanism is implemented inside the local computing node, handling the I/O requests issued by the local node. Our SSDUP+ is designed mainly for the I/O server-oriented burst buffer architecture. However, the core techniques in SSDUP+, including detection of random data, the traffic-aware pipeline mechanism and log-structure metadata management, can also be applied to the local burst buffer mechanism in computing nodes.

The details of these four components are explained next.

\subsection{Detecting Random Access}

Random accesses may cause the significant seek delay due to the need of moving the disk
head to the correct position before transmitting any data. The I/O behaviour of an HPC
application can be influenced by many factors, such as the way in which the processes access the data, the data size of each I/O request, the number of processes participating in I/O, the number of data servers, the implementation of the I/O subsystem in the Unix kernel. All these factors may lead to the random accesses. It is difficult for a single process to detect the nature of the randomness in data accesses.

In order to understand the relationship between the access pattern and disk latency, we carried out a series of experiments using IOR~\cite{IOR}, a benchmark to test the performance of parallel file systems under various conditions. Specifically, three
different access patterns were tested: \textit{segmented-contiguous},
\textit{segmented-random}, and \textit{strided}~\cite{bent2009plfs}. The total
data size tested was 16 GB. The size of each I/O request was 256 KB. The number of
processes varied from 4 to 128. 10 nodes were used in the experiments, 8 of which were compute nodes and 2 of which are I/O nodes. OrangeFS 2.9.3 were installed on both I/O nodes.
Other details of the experimental platform are given in Section 3.

\textbf{Segmented-Contiguous}: In this pattern, each process accesses $1/n$ portion
of the shared 16GB file ($n$ represents the number of processes). Each process
issues the sequential requests to access its part of data.

\textbf{Segmented-Random}: This pattern is similar to segmented-contiguous.
The only difference is that each process issues the random I/O requests.

\textbf{Strided}: In this pattern, the number of processes is $n$. In the $i$-th iteration, process with ID $j$ issues an I/O request to access the data at the offset $i * n + j$.

\begin{figure}[htb]
  \vspace{-0.2cm}
  \centering
  \includegraphics[width=0.4\textwidth, height=4.7cm]{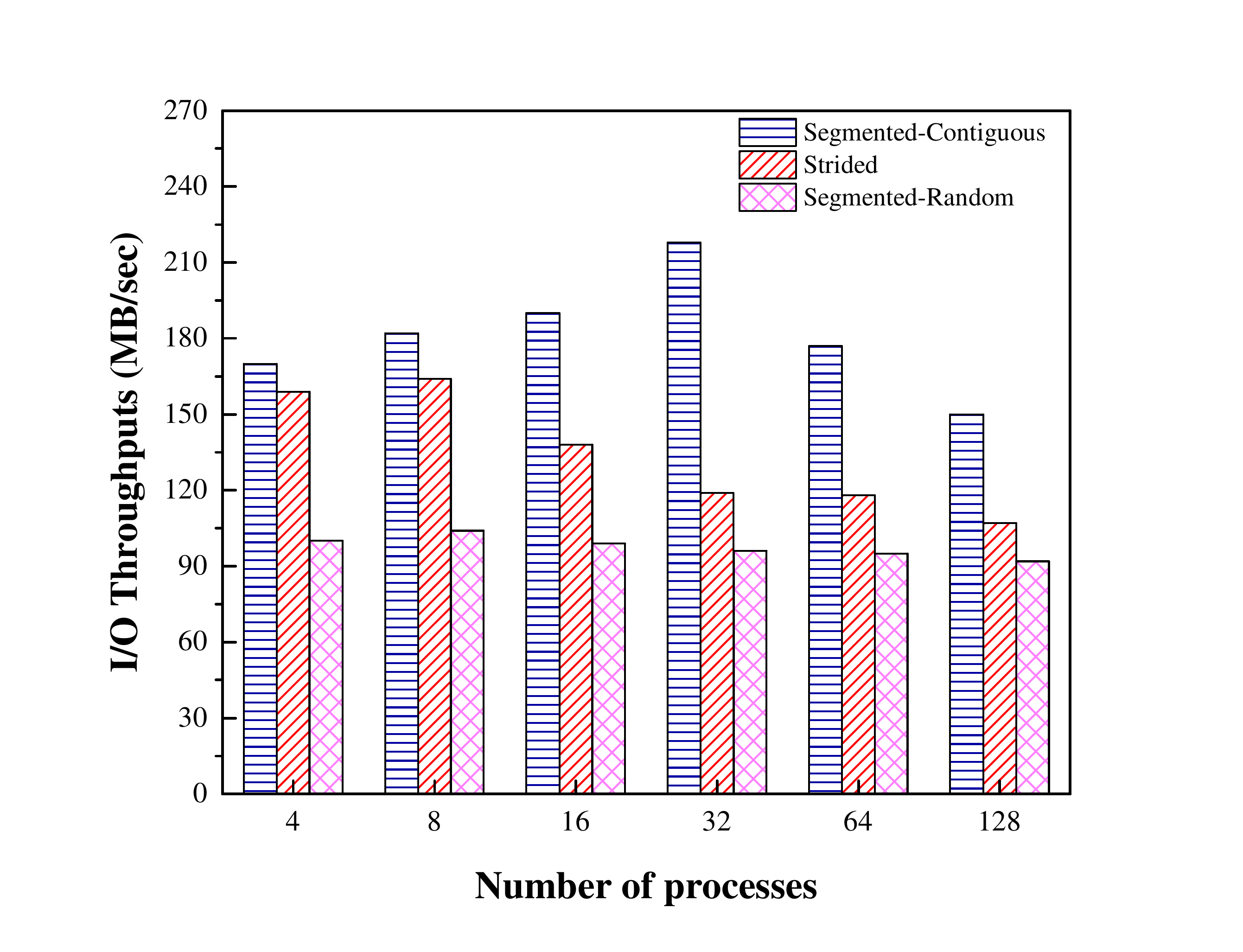}
  \vspace{-0.2cm}
  \caption{The I/O throughput of IOR with different access patterns and different number of processes}
  \label{figure-throuputs-design}
  \vspace{-0.2cm}
\end{figure}

The experimental results are presented in Figure~\ref{figure-throuputs-design}.
It can be observed that the throughput increases first and then drops
with the increase in the number of processes in both segmented-contiguous
and strided patterns. The reason for this result is because
OrangeFS uses \textit{asynchronous I/O} (AIO) as its default trove method~\cite{orangefs}.
AIO~\cite{aio} can improve its throughput when the number of I/O
participants grows. This is because the CFQ (\textit{Completely Fair Queuing})~\cite{cfq} scheduler tends to queue more requests so as to achieve better spatial locality by sorting and merging these requests.
When the number of processes increases further (from 16/32 processes to 128 processes),
the throughput of the segmented-contiguous accesses drops from 218 MB/s to 150 MB/s,
while the throughput of the strided accesses drops from 164 MB/s to 107 MB/s,
amounting to 31\% and 34\% degradation, respectively.
The reason for this is because the size of the queue in the CFQ scheduler is limited. When the number of I/O processes increases further, CFQ will not be able to achieve even higher locality.

The throughput of the segmented-random accesses remains at around 95 MB/s, because
the CFQ scheduler can hardly merge any requests when the offsets are non-consecutive.

\begin{figure*}[htb]
  \vspace{-0.2cm}
  \centering
  \subfloat[\textit{Offset distribution of segmented-contiguous accesses}]{
    \vspace{-0.2cm}
    \includegraphics[width=0.3\textwidth, height=4cm]{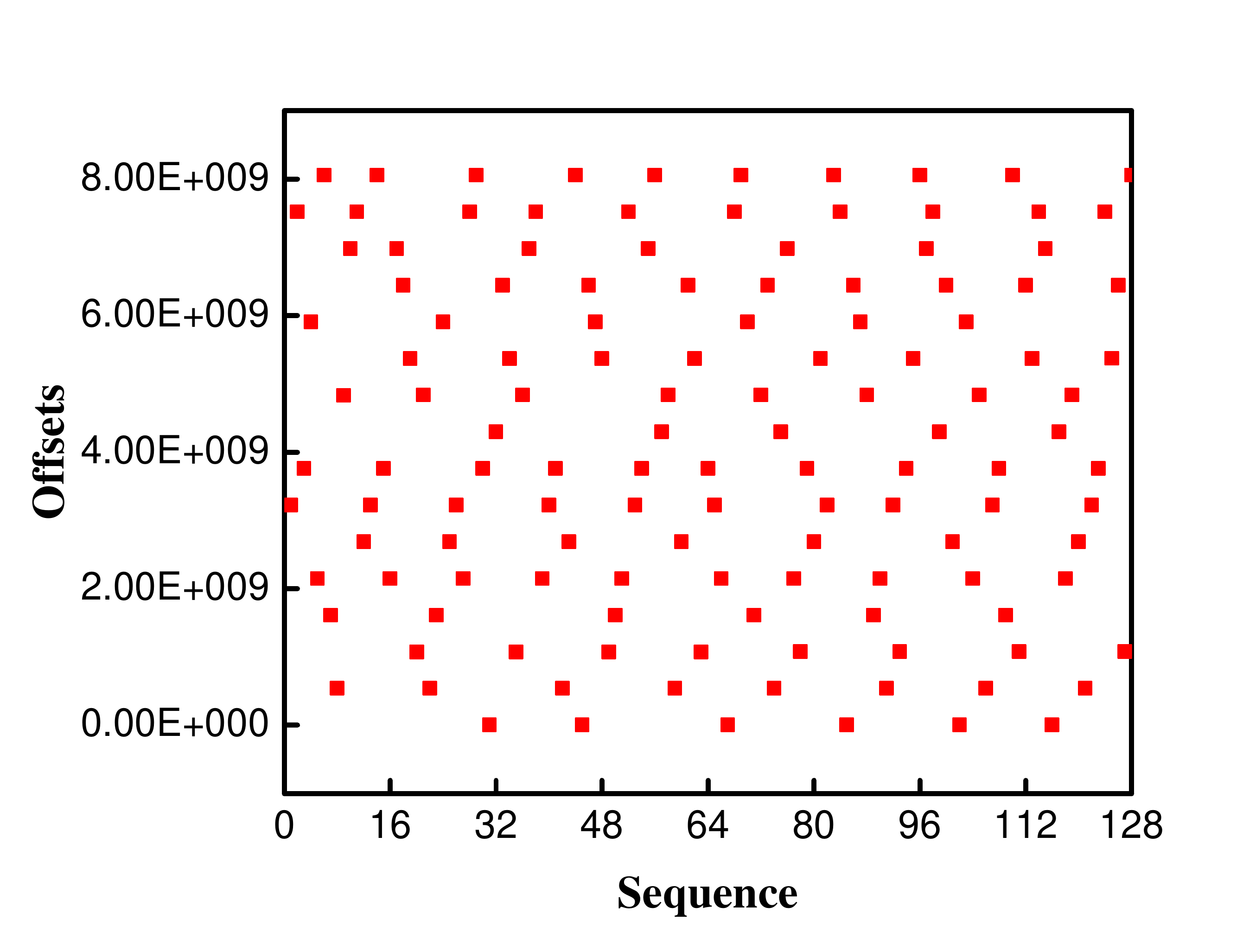}
    \label{conti-offset}}
  \hspace{0.01\linewidth}
  \subfloat[\textit{Offset distribution of segmented-random accesses}]{
    \includegraphics[width=0.3\textwidth, height=4cm]{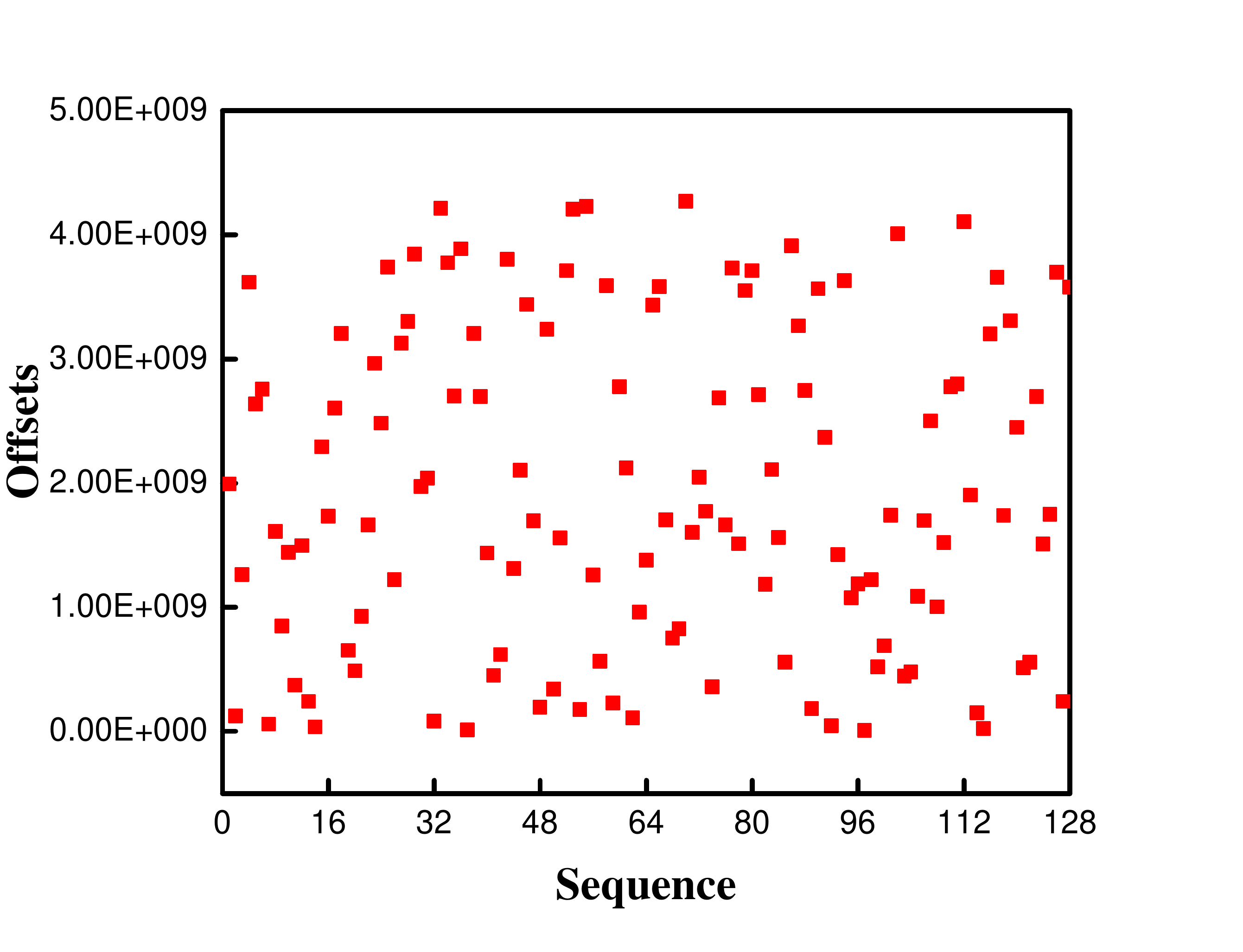}
    \label{random-offset}}
  \vfill
  \subfloat[\textit{Offset distribution of strided accesses}]{
    \includegraphics[width=0.3\textwidth, height=4cm]{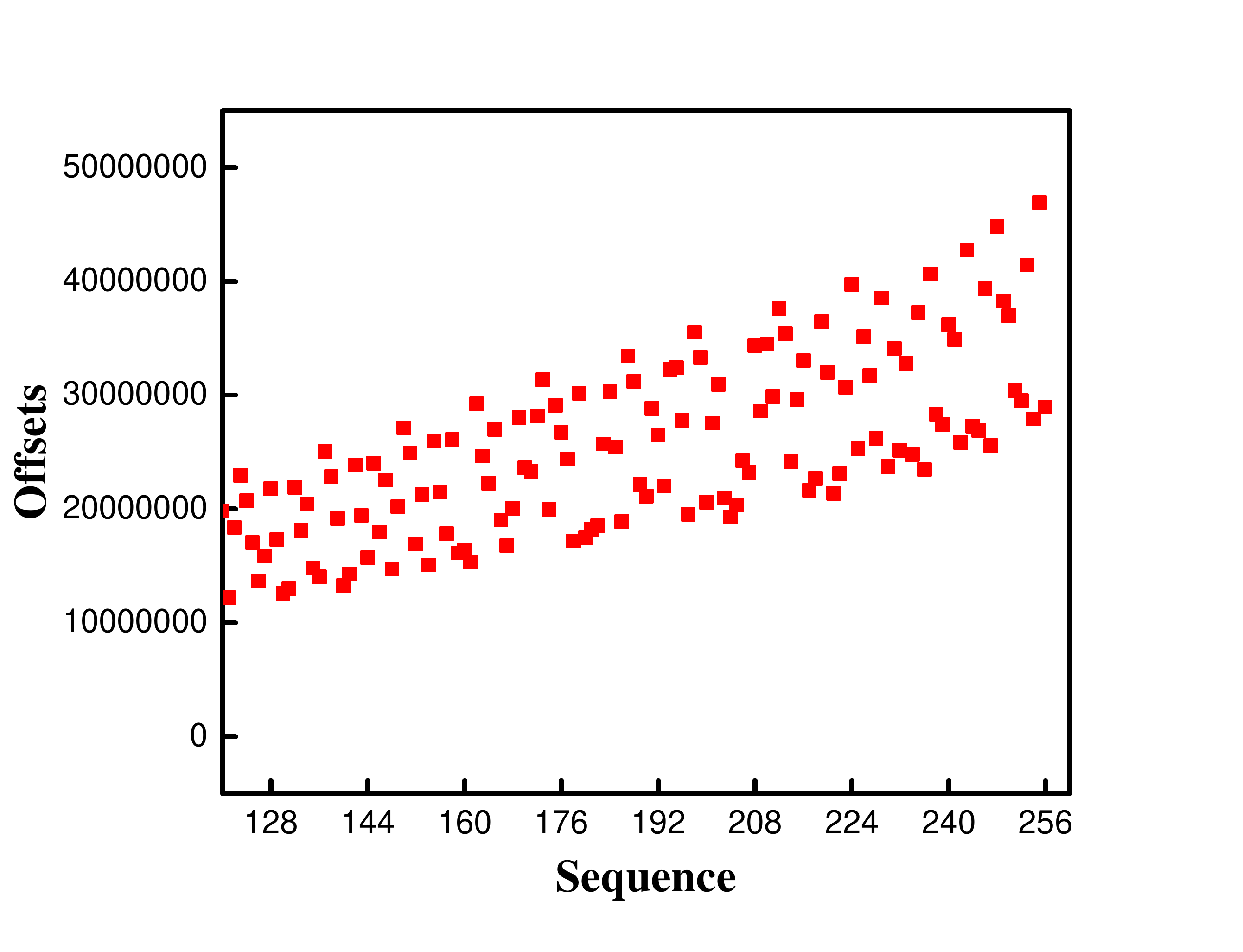}
    \label{strided-offset}}
  \hspace{0.01\linewidth}
  \subfloat[\textit{Offset distribution of mixed loads(segmented-contiguous$\times$segmented-rando)}]{
    \includegraphics[width=0.3\textwidth, height=4cm]{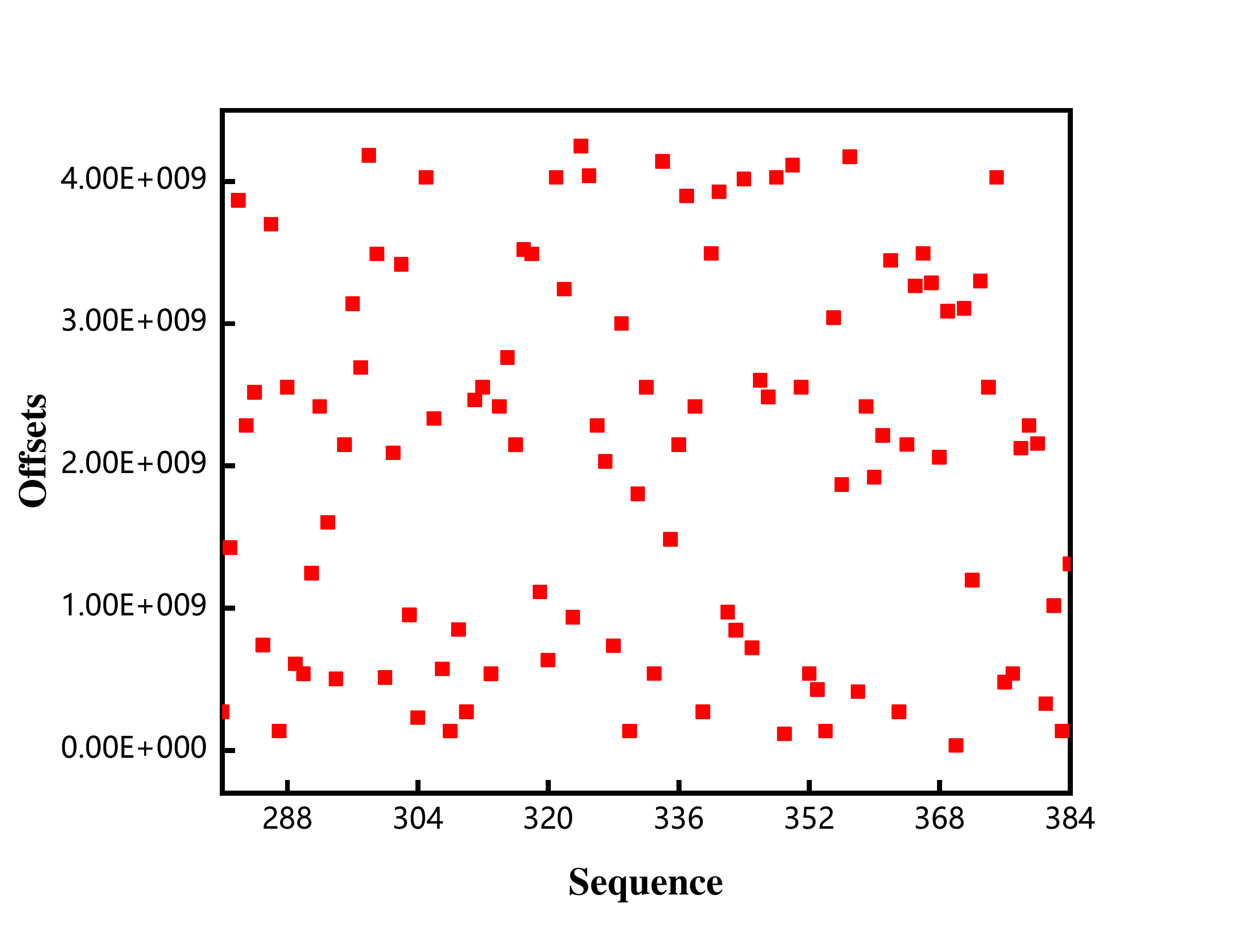}
    \label{mixed-loads-offset}}
  \caption{Offset distribution of various access patterns}
  \label{offsets}
  \vspace{-0.4cm}
\end{figure*}

To verify these observations, we also traced the offsets of these different access patterns. 
%Figure~\ref{offsets} shows the distribution of the offsets for the case where there are 16 processes. In this experiment, 65536 (16GB/256KB = 65536) requests were issued in total. Figure~\ref{conti-offset} depicts the offset distribution of the first 128 requests.
Figure~\ref{offsets} shows the offsets for a single application in three access pattern and the offsets for two applications which are segmented-contiguous $\times$ segmented-random mixed loads. In this experiment, 65536 (16GB/256KB = 65536) requests were issued by 16 processes in total for every single application, and 65536 (8GB/256KB+8GB/256KB = 65536) requests were issued by 32 processes in total for mixed loads.
Figure~\ref{conti-offset} depicts the offset distribution of the first 128 requests.
As shown in the figure, the offsets of the segmented-contiguous accesses are regular,
whereas the segmented-random accesses show the completely random offsets
(Figure~\ref{conti-offset} and Figure~\ref{random-offset}).
The offsets of the strided accesses (Figure~\ref{strided-offset}) are compact with slight fluctuations. Besides, the offsets shown in Figure~\ref{mixed-loads-offset} seem rather random.
\par
Similar to what we did for CFQ, We also used 128 requests as a unit block and sorted the offsets in every block of 128 requests.
After sorting the offsets, the offsets of the request blocks manifested a much better order.
If the offsets of the requests in a block are adjacent, they will be merged and the seek distance is regarded zero, although the offsets of the consecutive requests are not contiguous.
The sorted access order is beneficial because without the sorting, the disk head has to
move back and forth from the offset of one request to that of next consecutive request, which leads to the longer access delay.

We introduce a notion of \textit{random factor} (\textit{RF}) in SSDUP+ to indicate the number of disk head movements. After the offsets are sorted,
if the distance between their offsets
equals to the request size, we regard two requests as the sequential requests and assign the random factor to be 0, which represents the fact that the disk head does not need to seek the disk location for next request.
Otherwise, the requests are considered random and the random factor is 1, meaning that the disk head needs to move once. We calculate the sum of random factors, denoted by $S$, in every request stream, using Equation ~\ref{rf-sum}, where $N$ is the number of requests in a request stream.

\begin{equation} \label{rf-sum}
S = \sum_{i=1}^{N-1} RF_i \qquad
\end{equation}

\begin{figure}[htb]
  \vspace{-0.2cm}
  \centering
  \includegraphics[width=3.5in,height=3cm]{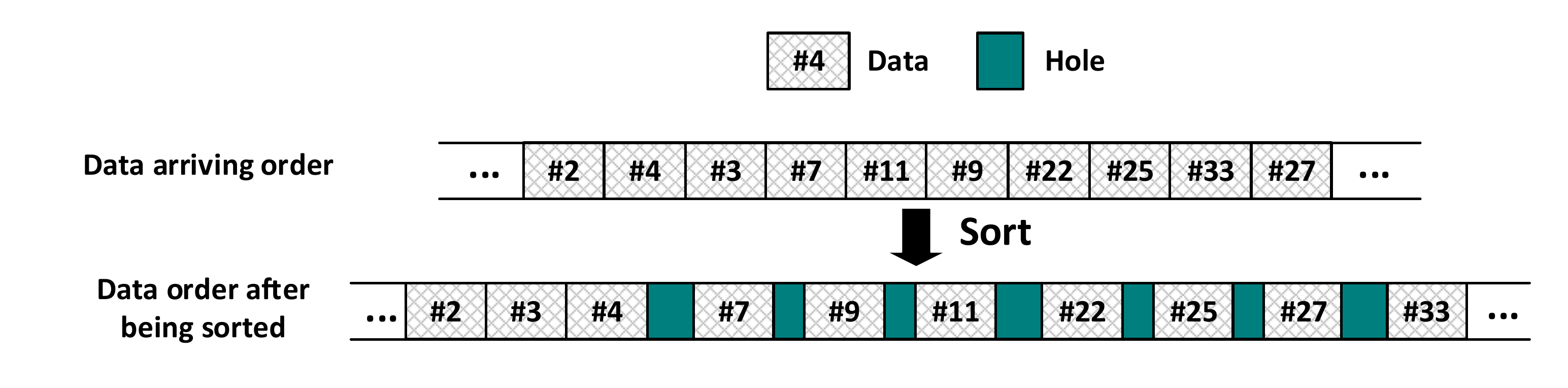}
  \vspace{-0.2cm}
  \caption{The sorting procedure in a requests stream}
  \label{rf}
  \vspace{-0.3cm}
\end{figure}

Figure~\ref{rf} illustrates the sorting procedure for a request stream. Requests
in a stream may arrive out of order because of the random or irregular
accesses issued by the processes and also the competitions among processes. As we have discussed above,
in this example, the random factor between requested data item \#2 and  \#3 is 0 because after sorting, the offset distance between these
two requests equals to the request size. On the other hand, the random factor
between data item \#4 and data item \#7 is 1.
Note that the random factor is defined based on logical address instead of
physical address. Although using the logical address to indicate the movement of disk head
is not 100\% accurate, the disk seek time is linearly related to the logical
address distance in most cases~\cite{huang2005fs2}. In addition, there is no need to distinguish between different applications when sorting the logical addresses of the requests. This is because in a single request stream, there is little correlation between the logical addresses of requests for different applications.

\begin{figure*}[htb]
  \vspace{-0.2cm}
  \centering
  \subfloat[\textit{Offset distribution of segmented-contiguous accesses after sorting}]{
    \vspace{-0.2cm}
    \includegraphics[width=0.3\textwidth, height=4cm]{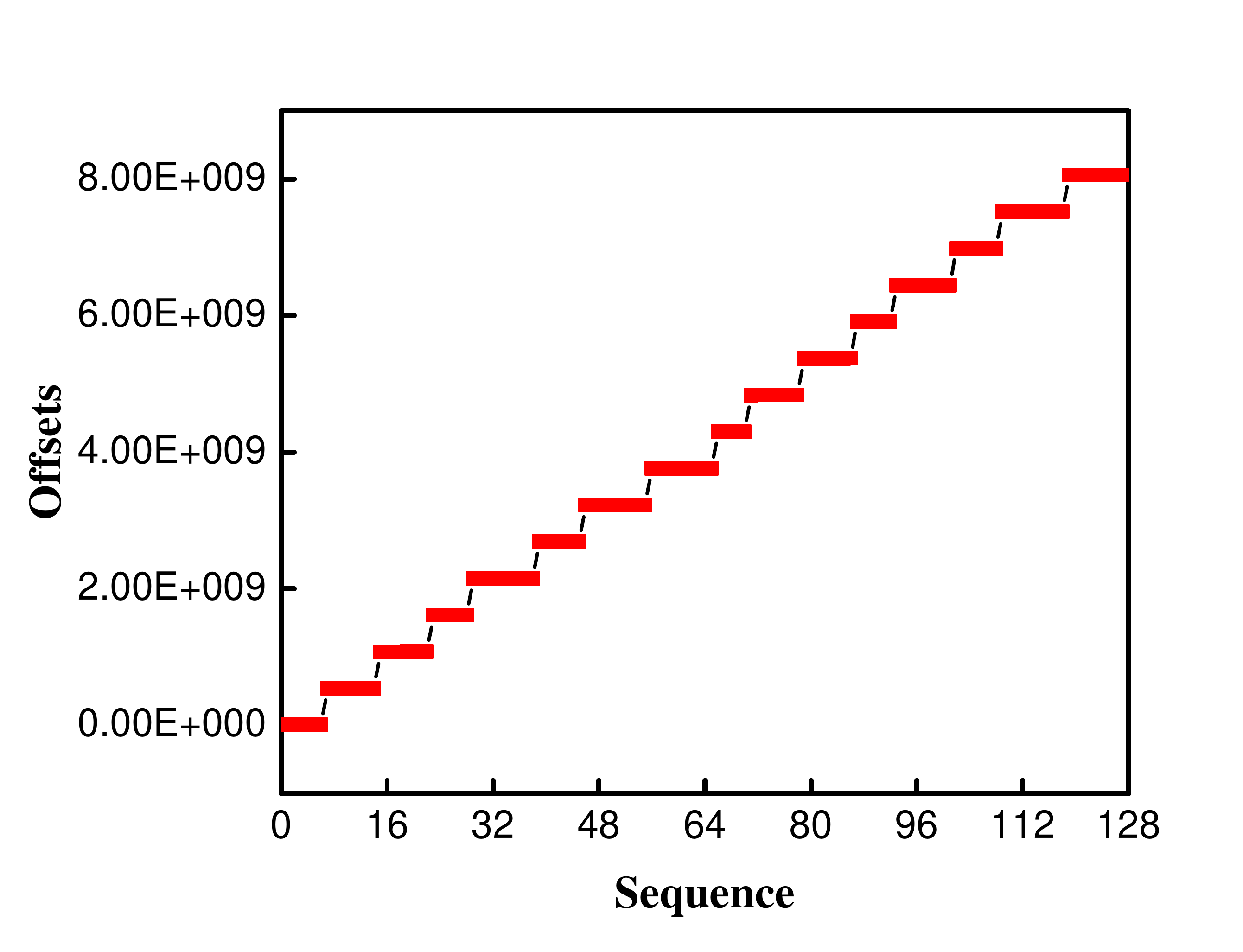}
    \label{conti-offset-sorted}}
  \hspace{0.01\linewidth}
  \subfloat[\textit{Offset distribution of segmented-random accesses after sorting}]{
    \includegraphics[width=0.3\textwidth, height=4cm]{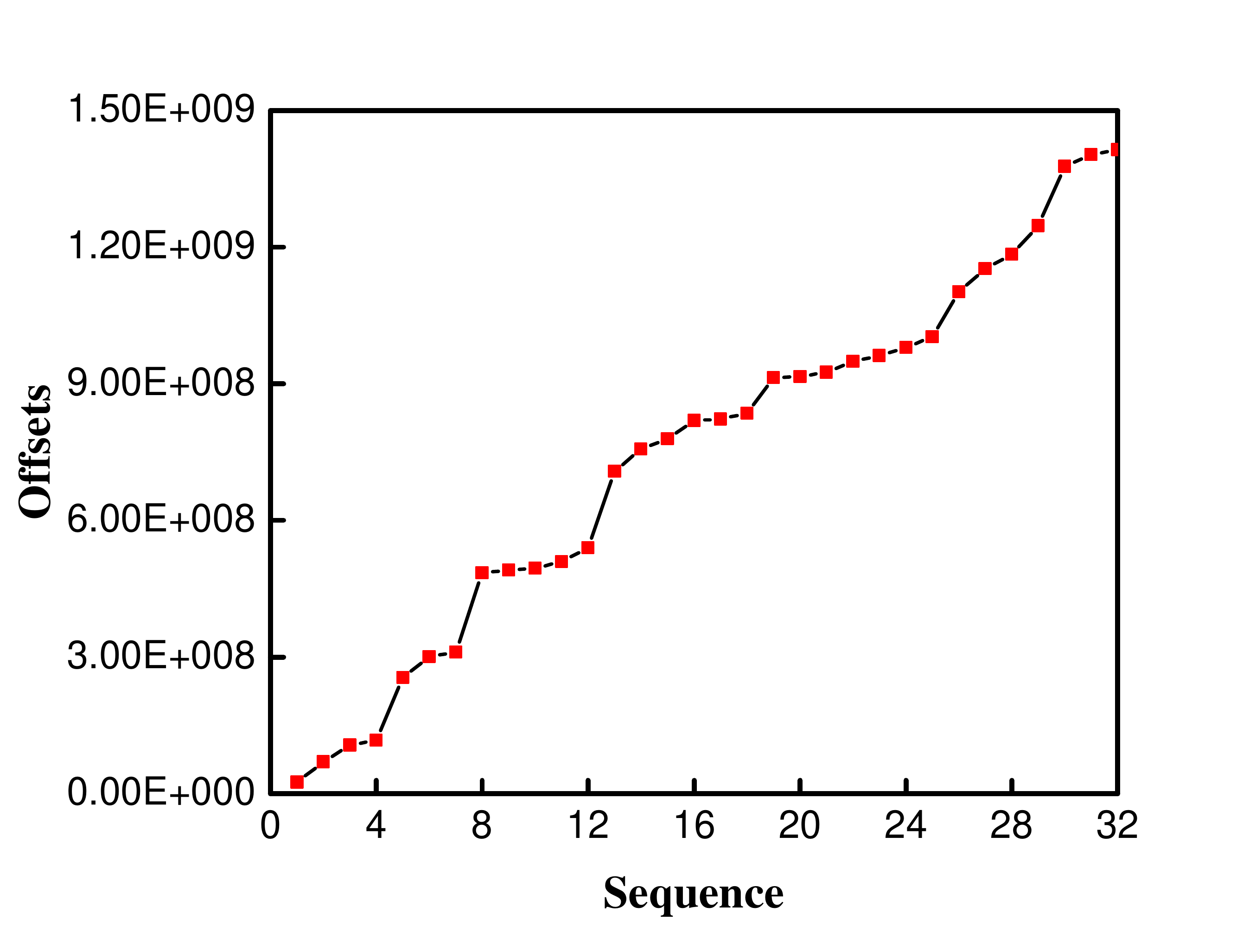}
    \label{random-offset-sorted}}
  \vfill
  \subfloat[\textit{Offset distribution of strided accesses after sorting}]{
    \includegraphics[width=0.3\textwidth, height=4cm]{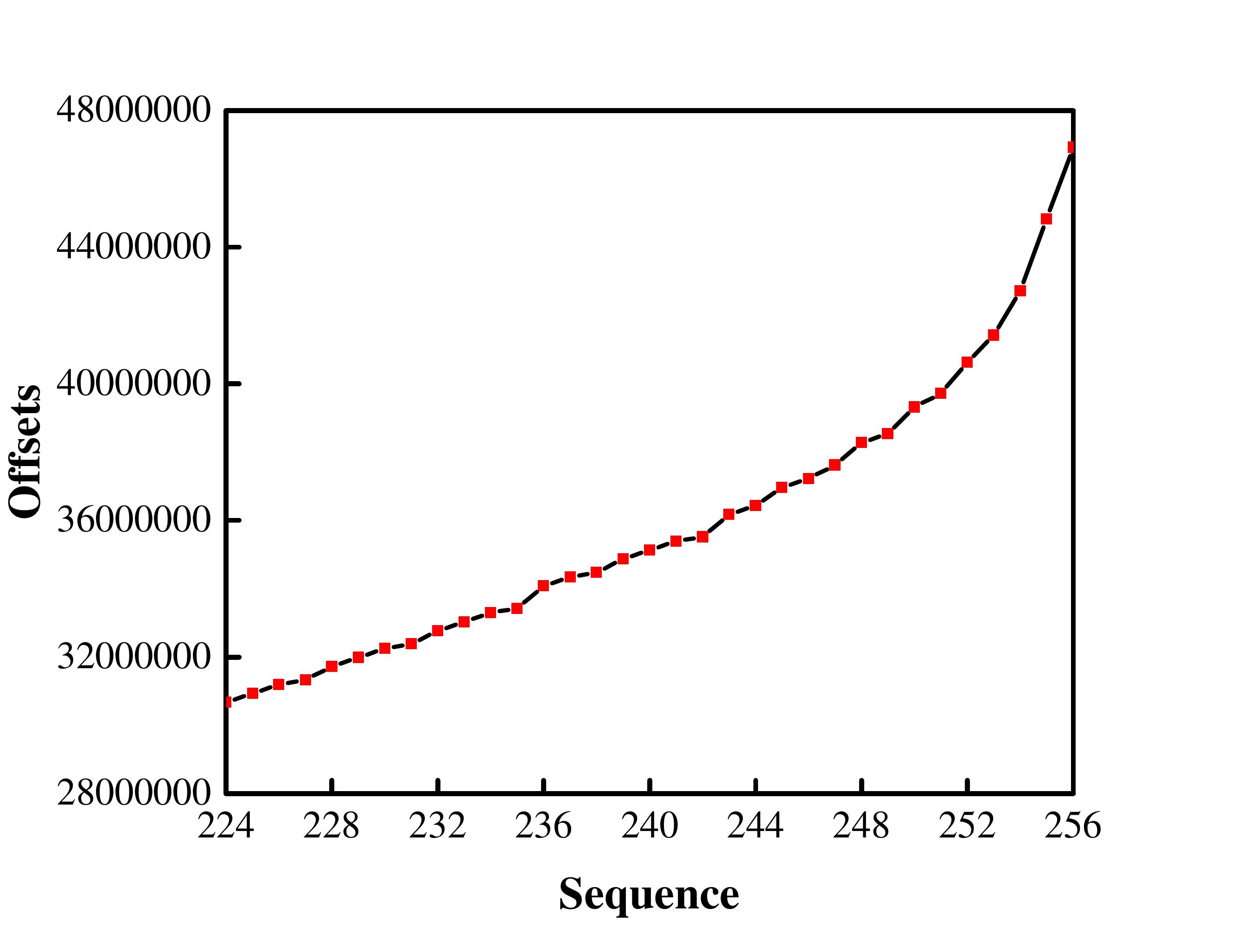}
    \label{strided-offset-sorted}}
  \hspace{0.01\linewidth}
  \subfloat[\textit{Offset distribution of mixed load after sorting}]{
    \includegraphics[width=0.3\textwidth, height=4cm]{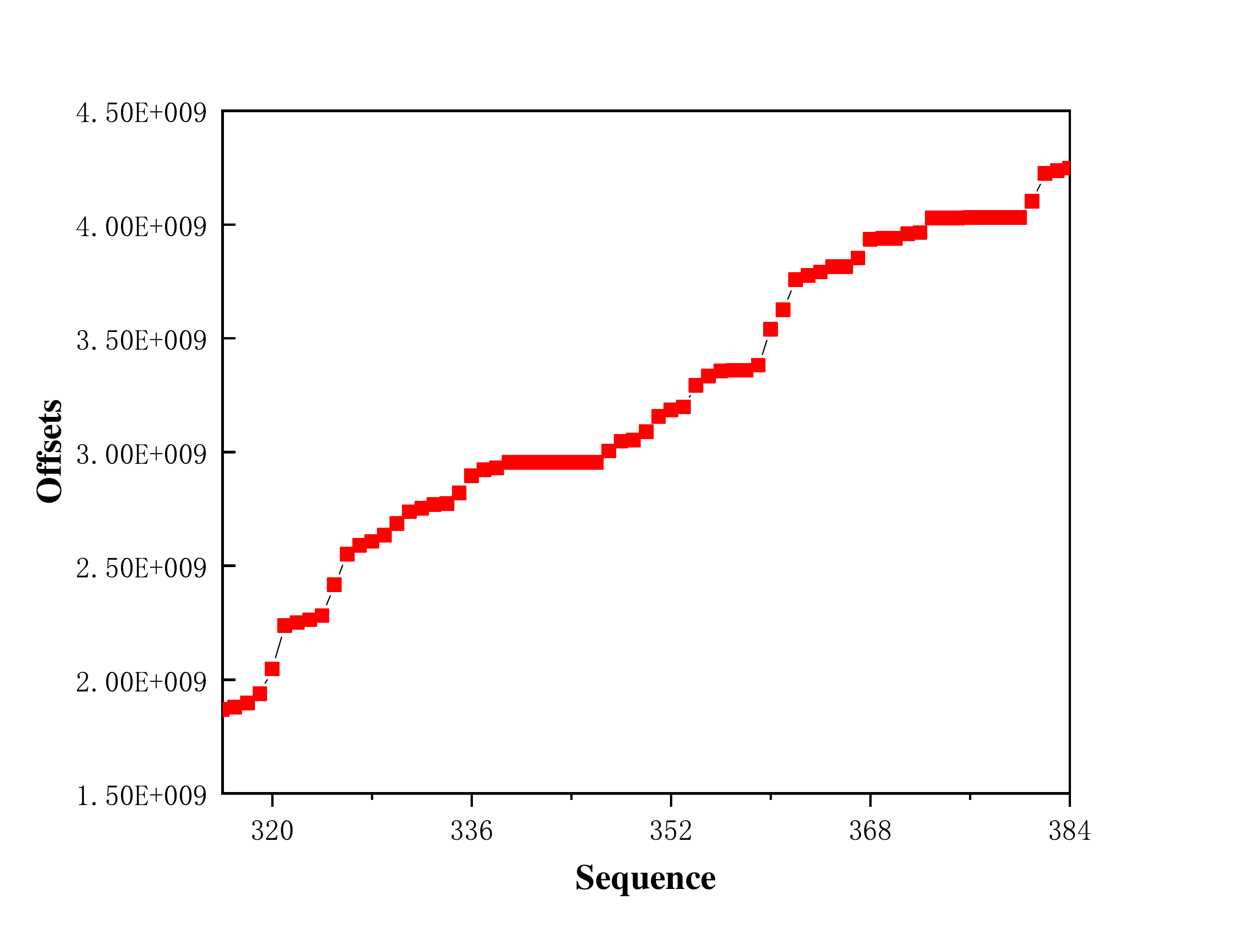}
    \label{mixed-offset-sorted}}
  \caption{Offset distribution of various access patterns with 16 processes after sorting}
  \label{offsets-sorted}
  \vspace{-0.3cm}
\end{figure*}

Figure~\ref{offsets-sorted} shows the distribution of offsets after the requests are sorted.
In these figures, a solid black line connecting two discrete red lines (or dots) represents a disk seeking movement. To make these figures more readable, we only draw 32
offsets out of 128 for both segmented-random and strided patterns. We draw 64 offsets out of 128 for the mixed load.

In a request stream, there are 128 requests and therefore the maximum number of disk seeking that the disk head has to perform when serving the request stream is 127 (i.e., 128-1). As can be seen from figure~\ref{offsets-sorted} (the random factor equals the total number of discrete lines (or dots) in the figure minus 1), the random factor of segmented-contiguous
accesses becomes 15 after sorting, which means that when serving the requests the disk head needs to move 15 times, accounting for 11\% of the maximum 127 movements. 

We introduce another notion, called \textit{random percentage}, to represent the level of randomness in a request stream. The random percentage in figure~\ref{offsets-sorted} is 11\%. The total random factor of the segmented-random accesses is 127, which means the random percentage is 100\% (i.e., there is no reduction of disk movement at all). In the strided accesses, the total random factor is 57 and the random percentage is 45\%.  

The situation becomes more complicated under the mixed load. As shown in Figure~\ref{mixed-offset-sorted}, some consecutive segments interleaved with random segments. The offsets of consecutive segments correspond to segmented-contiguous, while most offsets of random segments correspond to segmented-random. 
%most the front part denotes the requests issued by the IOR which is segmented-random mode, and the latter part denotes the requests issued by the IOR which is segmented-random mode. 
It can be observed that after sorting, the offsets show their respective characteristics when they are executed independently, which means that in the mixed load, the random factor exhibits a superimposed characteristic. The random percentage for the mixed loads is 71.88\% (the total random factor is 91).

\subsection{Data Redirection Based on Random Factor}

The data redirector component in SSDUP+ is designed to transmit the data to HDD or SSD. 
This component mainly consists of two parts, which calculate the random percentage in the incoming request stream and dynamically determine a threshold for the random percentage through the proposed adaptive algorithm. When the random percentage of the incoming request stream is higher than the threshold, the requests will be redirected to SSD. 
%Algorithm 1 outlines the data redirection process in the component.
\subsubsection{Calculating the Random Percentage}
When the execution of an application starts, the data is written to HDD. In the meantime,
the offsets of the requests in every request stream are traced and sorted. The random percentage, $percentage$, is calculated by $percentage = \frac{S}{N-1}$, where $S$ is the total random factor calculated by Eq. ~\ref{rf-sum} and $N$ is the number of requests in the request stream. The default length of a request stream is 128, which is the same as the queue size in the CFQ scheduler.
The length of a request stream can be re-configured when the CFQ queue size changes.

\begin{figure}[htb]
  \vspace{-0.1cm}
  \centering
  \includegraphics[width=0.40\textwidth, height=4.2cm]{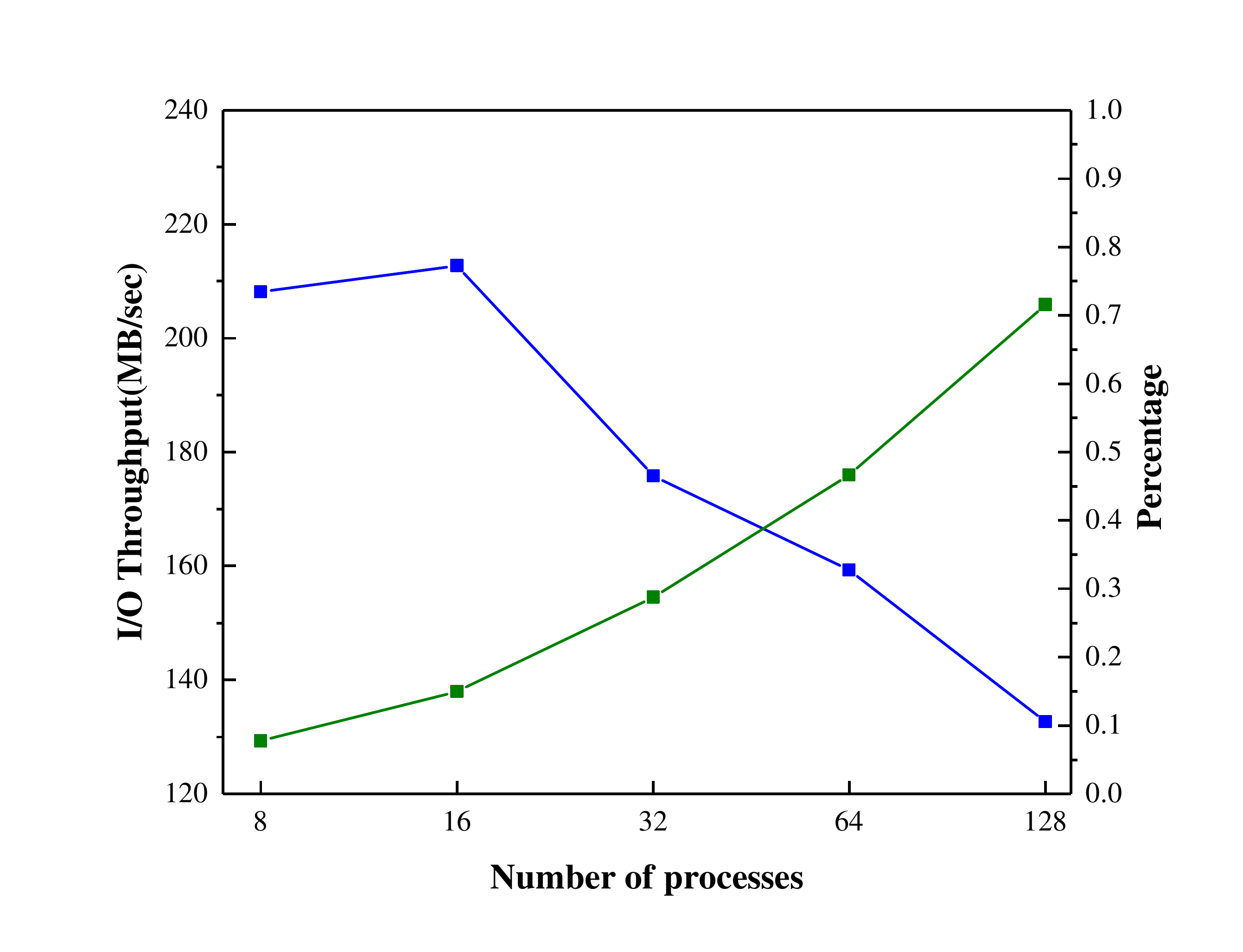}
  \vspace{-0.2cm}
  \caption{The changes in I/O throughput and random percentage of the request streams as the number of processes increases in OrangeFS}
  \label{percent-throughput}
  \vspace{-0.1cm}
\end{figure}

Typically, the greater random percentage of the request stream, the lower the I/O bandwidth is. To demonstrate this relationship, we use IOR to conduct a set of tests, in which the experimental environment is the same as that presented in Section 2.2. In this experiment, we calculate the random percentage of each workload whose access pattern is strided. The reason why we choose the strided pattern not others is because the changes in random percentage in segmented-contiguous and segmented-random are not prominent because of the features of their data accessing. The experimental results are presented in Figure~\ref{percent-throughput}. It can be observed that in the cases of 8, 16, 32, 64, 128 processes, the random percentages are 7\% 15\%, 28\%, 46\% and 71\%, respectively, while the I/O throughputs are 208.1MB/s, 211.76MB/s, 175.8MB/s, 159.29MB/s and 132.68MB/s, respectively. It is apparent that the I/O throughput decreases as the randomness of the workload increases.

\subsubsection{Determining the Threshold adaptively}
%In order to distinguish the random request more precisely,  we introduce a new parameter, $threshold$. If $percentage$ of the current request stream is greater than $threshold$, the requests are deemed random and the upcoming requests are redirected to SSD. If $percentage$ is less than $threshold$, the stream is regarded as sequential and the subsequent requests are redirected to HDDs.
In order to determine the random requests, we introduce a new parameter, the threshold of the random percentage, denoted by $threshold$. If $percentage$ of the current request stream is higher than $threshold$, the requests are deemed random and the incoming requests are redirected to SSD. If $percentage$ is lower than $threshold$, the stream is regarded as sequential and the subsequent requests are redirected to HDD.
In the previous version of this work, SSDUP, we set the "high-water mark" and "low-water mark" thresholds (more specifically, 45\% and 30\% in our prototype, respectively). When the $percentage$ is greater than the "high-water mark" threshold, the stream is considered random, and the upcoming requests are redirected into SSD. If the $percentage$ is less than a "low-water mark" threshold, the stream is treated as sequential and the subsequent requests are redirected to HDD.
\par
However, as the number of jobs which issue the I/O requests concurrently change over time, the I/O workload at the server side is dynamic. Simply setting a static, empirical threshold is not always accurate. Therefore, we propose a traffic-aware adaptive algorithm, which is able to dynamically adjust the threshold according to the current workload level.

The adaptive algorithm is designed to dynamically adjust $threshold$ as new request streams arrive. When a new request stream arrives, the $percentage$ of the stream is inserted into a list, called $PercentList$, in the increasing order. A $percentage$ is then selected from the $PercentList$ and used as the $threshold$. When the access pattern of the workload changes, the $PercentList$ will be emptied so that the access pattern of previous jobs does not interfere with the redirection of the request streams issued by new jobs. Equation \ref{thresholdgen} is used to calculate the element in $PercentList$ that is selected as $threshold$, where $avgper$ is the average over all elements in $PercentList$, as calculated in Equation \ref{avgpergen}. $avgper$ is used as the basis for selecting the element in $PercentList$. 

The rationale behind the selection of the elements from $PercentList$ is  explained as follows. When the values of recent random percentages are small, it indicates the randomness of the current workload is small and therefore, the element with a bigger index in $PercentList$ should be selected. Consequently, a less proportion of the incoming requests will be redirected to SSD. Otherwise, an element with a small index in $PercentList$ should be selected, so that the random percentage of more request streams will be higher than $threshold$ and therefore more requests will be redirected to SSD).

\begin{equation} \label{thresholdgen}
threshold = PercentList[(1-avgper)*(N-1)]
\end{equation}

\begin{equation} \label{avgpergen}
avgper = \frac{\sum_{i=0}^{N-1} PercentList[i] \qquad}{N}
\end{equation}

The adaptive algorithm that calculates the threshold is outlined in Algorithm~\ref{dataredirect}.

%\begin{algorithm}[t]
%  \SetAlgoNoLine
%  \caption{The adaptive algorithm for calculating threshold}
%  %\begin{algorithmic}[1]
%  \KwIn{The percentages of request streams}
%  \KwOut{$threshold$ for next request stream}
%  \Repeat{there is no incoming request stream}
%  {
%    \If{The access pattern has changed}{Clear $PercentageList$}
%    {
%      Insert {$percentage$} into {$PercentageList$} by in the ascending order \\
%      Calculate $avgper$ using Equation \ref{avgpergen}\\
%      Calculate {$threshold$} using Equation \ref{thresholdgen}
%    }
%  }
%  \label{dataredirect}
  %\end{algorithmic}
%\end{algorithm}

%

Note that $percentage$ is calculated based on the latest 128 requests that SSDUP+ has received and that the comparison between $percentage$ and $threshold$ is used to guide the direction of the upcoming (future) requests. This method is effective because many HPC applications manifest the stable access
pattern or smooth change in access pattern ~\cite{yin2013pattern, wang2003profile}. It is very rare that the access pattern of a HPC application changes abruptly. Moreover, the data redirector module and the redirection algorithm work by tracking and using the properties of the data (i.e., the offsets and the sizes of the requests), not the data itself. SSDUP+ does not change the data accessing behaviour in any way.

\begin{figure}[htb]
  \vspace{-0.1cm}
  \centering
  \includegraphics[width=0.40\textwidth, height=4.2cm]{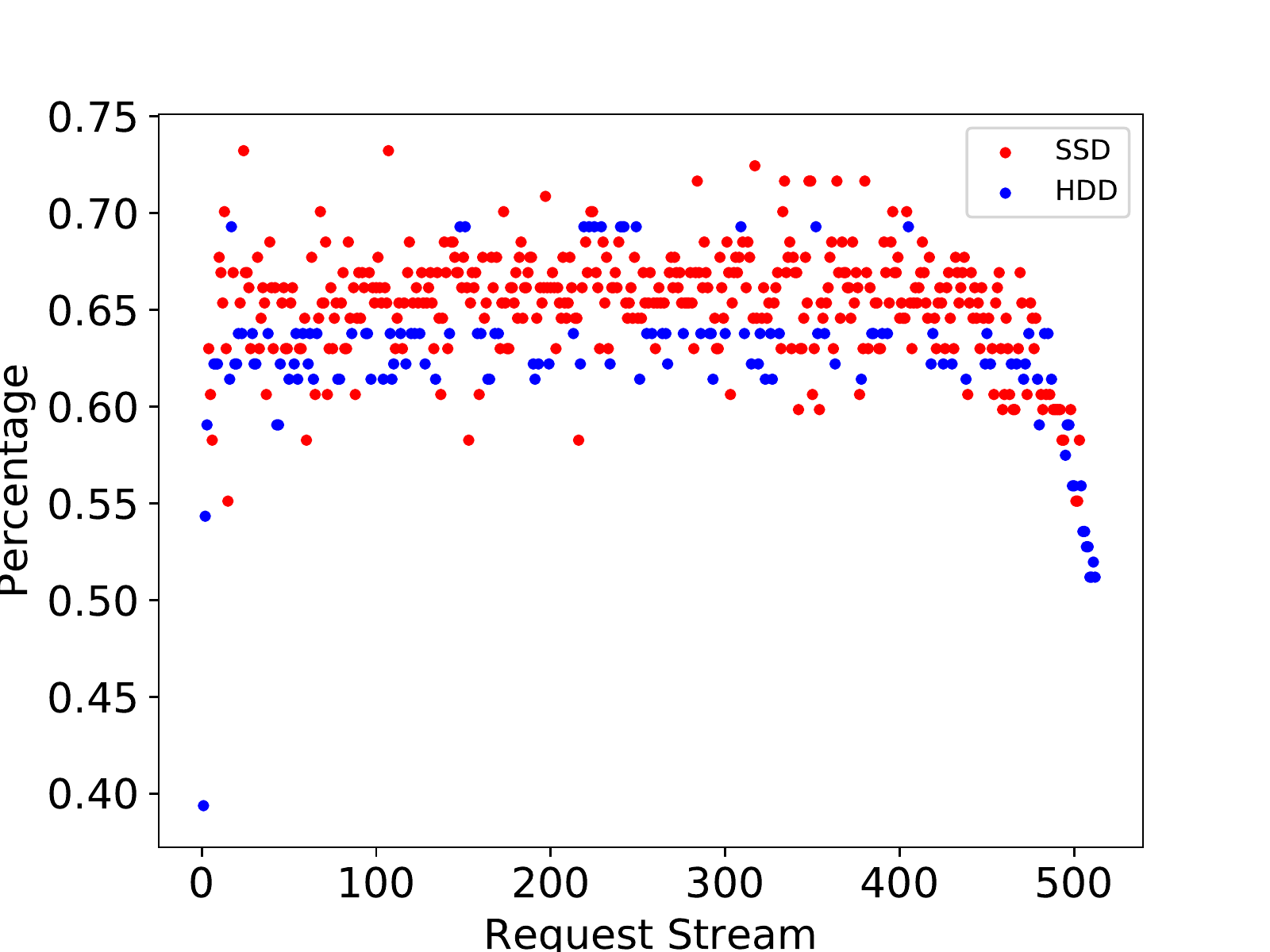}
  \vspace{-0.2cm}
  \caption{distribution of the percentages of the request streams}
  \label{algorithm2}
  \vspace{-0.1cm}
\end{figure}

We present the following case study to illustrate the adaptive determination of $threshold$. When we run the IOR instance, we record $Percentage$ of the latest 10 request streams in
$PercentList$. The set of percentages in the list is 0.3937, 0.5433, 0.5905, 0.6299, 0.6062, 0.5826, 0.622, 0.622, 0.622, 0.6771. The thresholds that we calculated are 0.5, 0.5433, 0.5433, 0.5433, 0.5905, 0.5826, 0.5826, 0.5905, 0.5905, 0.6062 as the request streams arrive. It is clear that the thresholds vary from the random percentages of the request streams. The request streams that are directed to SSD are those with the percentages of 0.6299, 0.6062, 0.5826, 0.622, 0.622, 0.6771. The decision of directing the data to SSD is deemed a correct decision when the percentage of the request stream is greater than the average threshold. In this case study, the IOR instance contains 512 request streams. The proportion of successful directions is 79.48\%. As Figure~\ref{algorithm2} shows, the red and blue circles represent the request streams directed to SSD and HDD, respectively. 
It is clear that more request streams with higher percentages are directed to SSD.

\par

In SSDUP+ there is no need to set the thresholds manually thanks to our traffic-aware adaptive algorithm. Comparing with SSDUP, whose thresholds take the empirical values, the SSD utilization can be further improved (i.e., less SSD capacity is required for achieving the same level of I/O performance).

\begin{figure}[htb]
  \vspace{-0.1cm}
  \centering
  \includegraphics[width=0.40\textwidth, height=4.2cm]{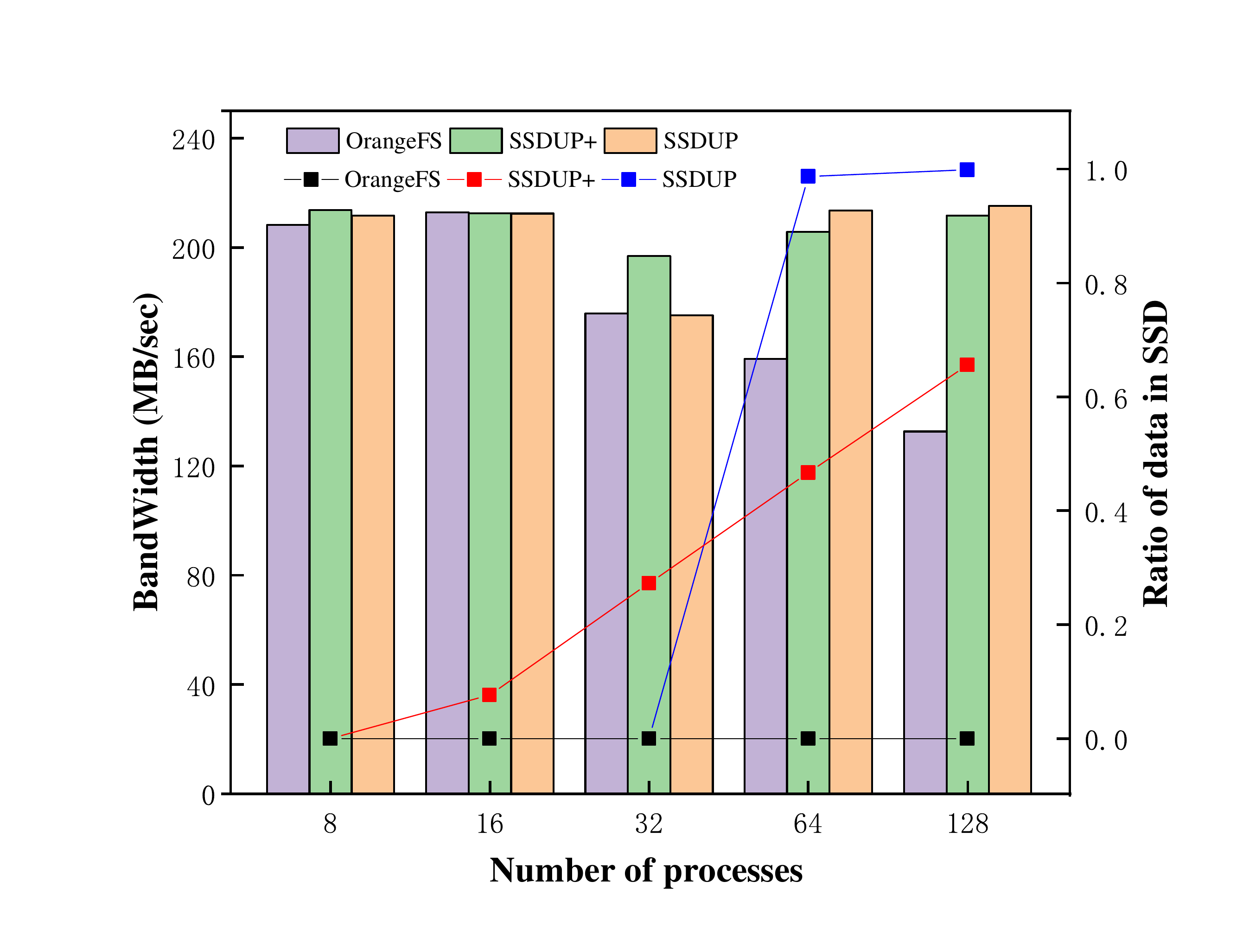}
  \vspace{-0.2cm}
  \caption{The changes in I/O throughput (the y-axis on the left) and ratio of data directed to SSD (the y-axis on the right) as the number of processes increases}
  \label{comparisons}
  \vspace{-0.1cm}
\end{figure}

To better demonstrate the advantage of our traffice-aware adaptive algorithm, we conducted a set of micro benchmark to compare the performance in terms of throughput among SSDUP+, native OrangeFS and SSDUP. IOR is still used in the experiments. The total data size is 16GB and the access pattern is strided. The number of processes varies from 8 to 128. Figure~\ref{percent-throughput} shows that as the number of processes increases, the I/O throughput achieved by OrangeFS gradually decreases due to the contention between multiple processes. At the same time, the random percentage increases. These trends show that the throughput and the random percentage manifest a good inverse correlation within a certain range. Namely, when we observe the increase in random percentage, it is very likely that the throughput is decreasing. 
%As shown in Figure~\ref{comparisons}, the red squares and triangles represent the throughput of SSDUP and OrangeFS, respectively. The blue-squares and the blue-triangle curves represent the SSD usage (the y-axis on the right) of SSDUP and OrangeFS respectively. %Ligang: It is better to change the colors of the curves in figure 8 as we discussed or change the figure legend to indicate which curve uses which y-axis.
In addition, figure~\ref{comparisons} shows the difference in throughput among SSDUP+, OrangeFS and SSDUP with different numbers of processes. When the number of processes is 8 and 16, all filesystems show excellent performance because of the low randomness (OrangeFS: 208.1MB/s with 8 processes, 212.76MB/s with 16 processes; SSDUP+: 213.6MB/s with 8 processes, 212.48MB/s with 16 processes; SSDUP: 211.67MB/s with 8 processes, 212.38MB/s with 16 processes). When there are 32 processes, the throughputs of SSDUP+, OrangeFS and SSDUP decrease to 196.9MB/s, 175.8MB/s and 175.06MB/s. It can be seen that the  native OrangeFS and SSDUP show much greater decrease than SSDUP+. This is because that in SSDUP+, a portion of data (about 27.25\% in Figure 7) is identified as random requests based on the adaptive algorithm and is directed to SSD, while SSDUP cannot redirect the data accurately due to the manual setting of the threshold. When the number of processes further increases to 64, the throughput of the native OrangeFS decreases further (to 159.29MB/s) while the throughputs of SSDUP+ and SSDUP remain high since 46.68\% and 98.73\% of requests are directed to SSD respectively. Comparing with SSDUP, SSDUP+ saves more than 50\% of the SSD storage space in this case without sacrificing the performance. When there are 128 processes, the throughput of the native OrangeFS drops to 132.68MB/s,  while SSDUP+ and SSDUP still retain high throughputs. The proportions of requests that are directed to SSD are 65.63\% and 99.9\% in SSDUP+ and SSDUP, respectively. From this figure, we can see that with our adaptive algorithm, a good positive linear correlation is maintained between the amount of data directed to SSD and the randomness of the I/O load, which is the key to retain a high throughput in the system. In general, by using the adaptive algorithm in SSDUP+, random requests can be identified more accurately, saving more SSD space without sacrificing the throughput.

%

%\begin{equation}
%percentage = \frac{S}{N}
%\end{equation}

%\SetAlFnt{\small}
\begin{algorithm}[t]
  %\SetAlgoNoLine
  \caption{Redirection Algorithm}
  %\begin{algorithmic}[1]
  \KwIn{$requests$}
  %\KwOut{}
  Send {$requests$} to HDDs \\
  \Repeat {No incoming {$reqs$}}
  {
    Group $requests$ into requests stream \\
    Sort the offsets of $requests$ in requests stream\\
    Calculate $S$ (sum of $RF$)\\
    Calculate $percentage$ of $RF$\\
    Insert {$percentage$} into $PercentageList$\\
    \If { $percentage > threshold $ \textbf{and} current $requests$ being sent to HDDs}
      {Send $reqs$ of next requests stream to SSDs}
    \eIf{$percentage < threshold $ \textbf{and} current $requests$ being sent to SSDs}
      {Send $requests$ of next requests stream to HDDs}
      {Send $requests$ of next requests stream to the current device}
    }
  \label{dataredirect}
  %\end{algorithmic}
\end{algorithm}

\subsection{The Pipeline Scheme}
\subsubsection{Performance Improvement with the Pipeline scheme}
 %we have two options to process the data in \textit(R1), refreshing immediatly or delay.
In SSDUP+, we divide the total area of SSD into two equal-sized regions. Both regions are empty initially. When an I/O request is forwarded to SSD, SSDUP+
first selects an empty region, called \textit{Region1} (\textit{R1}), to buffer the incoming data.
When \textit{R1} is filled up and a new request is forwarded to SSD,
SSDUP+ then picks the other free region, called \textit{Region2} (\textit{R2}), to buffer the data. At the same time, \textit{R1} starts to flush its data to HDD. Namely, SSDUP+ flushes the data in \textit{R1} and writes the data to \textit{R2} simultaneously. When \textit{R2} is
filled up, the space of \textit{R1} has already been released. With the pipeline scheme, data flushing overlaps data writing. Consequently, the I/O throughput can be improved. From another perspective, it is more likely that there is the free SSD space when the requests arrive. Therefore, we can achieve the same I/O throughput with a smaller SSD capacity. We will conduct the detailed analysis about the effectiveness of the pipeline scheme in next subsection. 

In some occasions, both regions may be filled up (when the amount of random writes is larger than the SSD capacity). In this case, the system waits until a region becomes empty (the data in that region has been flushed). %Fig. \ref{pipeline} illustrates the two-region pipeline mechanism.
%
%\begin{figure}[htb]
% \vspace{-0.2cm}
% \centering
% \includegraphics[width=3.5in,height=3cm]{pipeline-region.pdf}
% \vspace{-0.4cm}
% \caption{SSDUP pipeline design with double buffer}
% \label{pipeline}
% \vspace{-0.3cm}
%\end{figure}

\subsubsection{Further Optimization with the Traffic-aware Pipeline strategy}
\label{subsec:furtheroptimization}
As the data is being flushed from SSD to HDD in the pipeline scheme, the data from other jobs may be written to HDD at the same time. The HDD writing from these two sources (i.e., data flushing from SSD flushing and data writing from jobs) may interfere each other, causing more disk seeking movements and consequently degrading the I/O throughput. To further optimize the I/O throughput, we propose a traffic-aware flushing strategy in the pipeline scheme, which judiciously flushes the data to reduce the I/O interference discussed above. 

The traffic-aware flushing strategy is designed based on the following reasoning. We can determine the randomness of I/O traffic by random percentage. When the data needs to be flushed, the traffic-aware flushing strategy checks the random percentage of current I/O traffic. If its random percentage is high, it suggests most requests are directed to SSD and consequently the traffic directed to HDD is low. Therefore, the data is flushed as normal. On the contrary, if the random percentage is low, it indicates that the traffic directed to HDD is high. Therefore, data flushing is paused to avoid the I/O interference. Data flushing stays paused until the random percentage becomes high again. Another advantage of this strategy is that we can make the data flushing phase to overlap the computing phase more so as to hide the flushing overhead. 

\subsubsection{Effectiveness Analysis of the Pipeline Scheme}
\par
%Ligang: add the analysis in SSDUP
Assume the I/O phase is divided into $n$ stages, and the size of data transmission in each stage is the same. $T_{HDD}$ and $T_{SSD}$ denote the time of writing one stage of data to HDD and SSD, respectively. Also assume that the entire capacity of the SSD is sufficient to accommodate $m$ stages of data ($m < n$). The total I/O time without the pipeline mechanism, denoted by $T_1$ , can be calculated as Equation ~\ref{t1-ori}. 

\begin{equation} \label{t1-ori}
T_1 =m * T_{SSD} + (n-m) * T_{HDD}
\end{equation}

In the pipeline mechanism, the SSD is divided into two equal-sized regions. Each region can handle $m/2$ I/O stages. So except the first and final $m/2$ stages, all other stages (out of \textit{n} stages) are handled in pipeline (i.e., while one region is flushing a stage of data, the other region is buffering the next stage of data). Therefore, there are in total $n-2 \times m/2$ stages that are handled in pipeline.  

Assume the first $m/2$ stages of data are written to the SSD region \textit{R1}. While the next $m/2$ stage of data are written to the region \textit{R2}, the data in \textit{R1} is flushed to HDD (i.e., these stages of data are handled in pipeline). Except the first and last $m/2$ stages, all other stages are fully handled in pipeline. Thus, the total number of I/O stages handled by the pipeline is $( n- 2 * m/2 )$. When the stages are handled in pipeline, the time spent by one stage is determined by the maximum between the time of flushing one stage of data (denoted by $T_{f}$) and the time of buffering one stage data (denoted by $T_{b}$). Therefore, the I/O time spent in writing $n$ stages of data with the pipeline mechanism, denoted by $T_2$, can be calculated by Equation ~\ref{t2-ori}.

\begin{equation} \label{t2-ori}
\begin{aligned}
T_2 &= \frac{m}{2} * T_{SSD} + (n - 2 * \frac{m}{2} ) * max\left \{ T_{f},T_{b} \right\} +\frac{m}{2} * T_{SSD} \\
&= m * T_{SSD} + (n - m ) * max\left \{ T_{f},T_{b} \right\}
\end{aligned}
\end{equation}

In Equation ~\ref{t2-ori}, $T_{b}$ is actually $T_{SSD}$ in Equation \ref{t1-ori}. As writing the data to SSD is faster than to HDD typically, $T_{f}$ is larger than $T_{b}$. So $T_2$ becomes:

\begin{equation} \label{t2-ori-1}
T_2 \approx m * T_{SSD} + (n-m) * T_{f} 
\end{equation}

In the flushing stage, the data is written back to HDD in a well-ordered fashion. Therefore, $T_{f}$ is the time of writing a stage of data to HDD sequentially. $T_{HDD}$ is the time of writing the data without ordering to HDD, which is typically larger than $T_{f}$. Therefore, $T_1$ is larger than $T_2$, which shows the pipeline mechanism reduces the I/O time of writing the data. 

Equation \ref{t2-ori-1} effectively models the I/O time spent in writing the data generated by a single application. In Equation \ref{t2-ori}, $T_f$ in the second term is the time for flushing one stage of data in SSD. However, as discussed in Subsection \ref{subsec:furtheroptimization}, as the number of I/O applications increases, there may be I/O interference when the data is flushed to HDD. Therefore, the flush time $T_f$ will increase. $T'_f$ denotes the time of flushing one stage of data under I/O interference. We have $T'_f > T_f$. So the total I/O time under the I/O interference (denoted by $T'_2$) is calculated by Equation \ref{t2-ori-2-prime}, which is larger than $T_2$. This shows that the I/O interference reduces the I/O performance, which is the reason why a traffic-aware pipeline strategy is designed in SSDUP+ to avoid the performance degradation caused by the I/O interference.

\begin{equation} \label{t2-ori-2-prime}
T'_2 \approx m * T_{SSD} + (n-m) * T'_{f} 
\end{equation}

\par
We designed a micro-benchmark to show the benefits of the traffic-aware flushing strategy. We set the capacity of a region of SSD as 4GB. We run two IOR instances concurrently: one with the segmented-contiguous pattern ($IOR_1$) and the other with the segmented-random pattern ($IOR_2$). The amount of data written by each IOR instance is 8GB, and the request size is 256KB. Note that the size of SSD is smaller than the amount of data generated by the applications in the setting, which is to simulate the situation which occurs typically in the HPC system and is therefore the reason why the Burst Buffer is deployed.
%The reason why we set the size of one region of SSD as 2GB and the total amount of data generated by applications as 16GB is that 
\par
\begin{figure}[htb]
  \vspace{-0.1cm}
  \centering
  \includegraphics[width=0.4\textwidth, height=4.2cm]{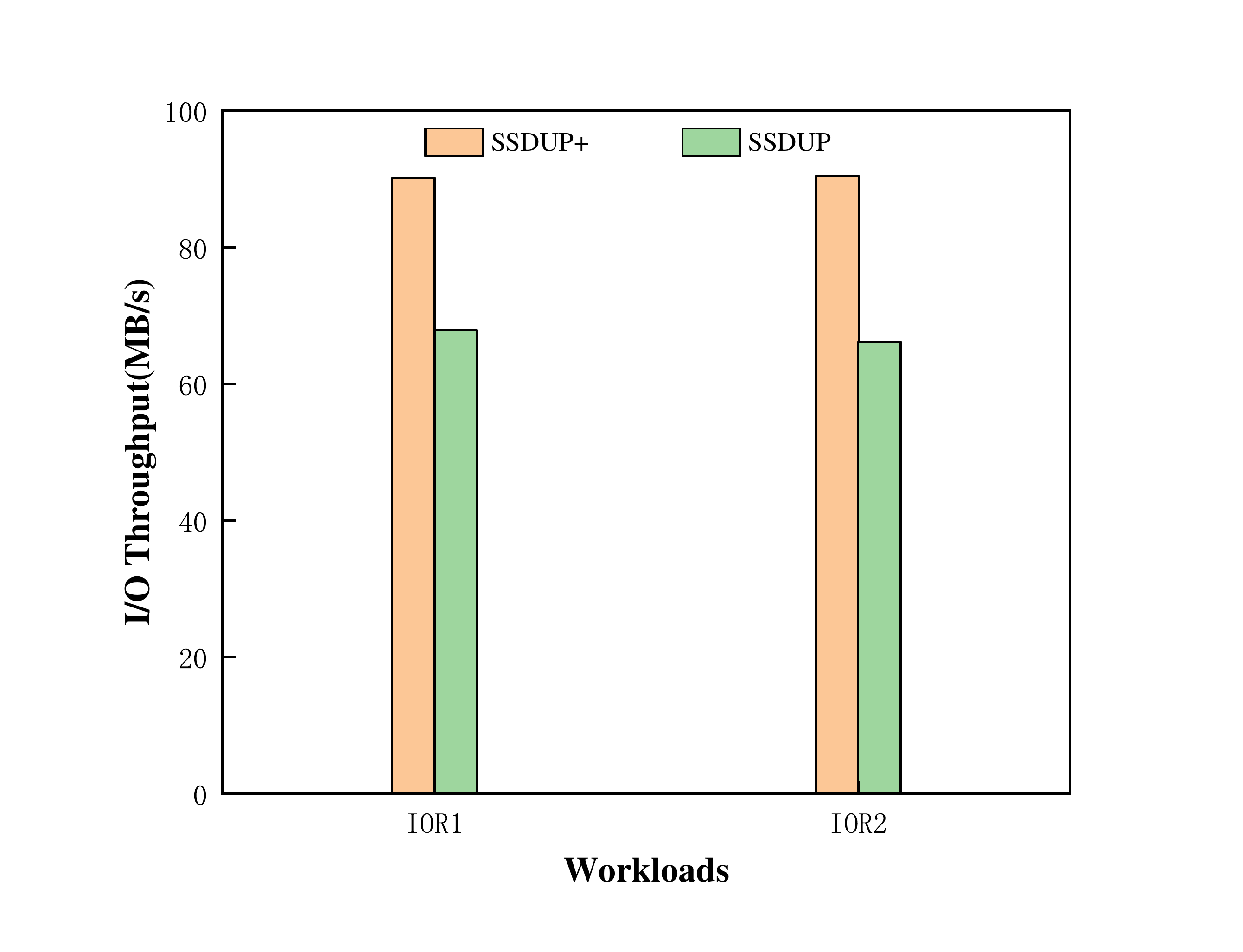}
  \vspace{-0.2cm}
  \caption{Comparison of I/O throughput between SSDUP+ and SSDUP}
  \label{flush-microbench}
  \vspace{-0.1cm}
\end{figure}

As shown in Figure~\ref{flush-microbench}, the I/O throughput of $IOR_1$ and $IOR_2$ are 90.21MB/s and 90.48MB/s respectively in SSDUP+, while they are only 67.84MB/s and 66.15MB/s respectively in SSDUP. The overall performance of SSDUP+ is 34.85\% higher than SSDUP. This is because under this mixed load, SSDUP+ uses a traffic-aware flushing strategy. In this benchmarking, SSDUP+ wrote about 10GB of data into SSD. Therefore, a total of three flush operations are triggered. The first and the second flush operation are paused in total by 17s and 19s, respectively, in order to avoid the interference in HDD between the data flushing operation and the data writing by the IOR instances. Because of the same reason, the third flush operation is paused until the IOR instances have completed the writing. 

\subsection{Buffer Management Using the AVL Tree}
\begin{figure*}[!t]
  \vspace{-0.2cm}
  \centering
  %\subfigure{
  \vspace{-0.2cm}
  \includegraphics[width=0.6\textwidth, height=3.5cm]{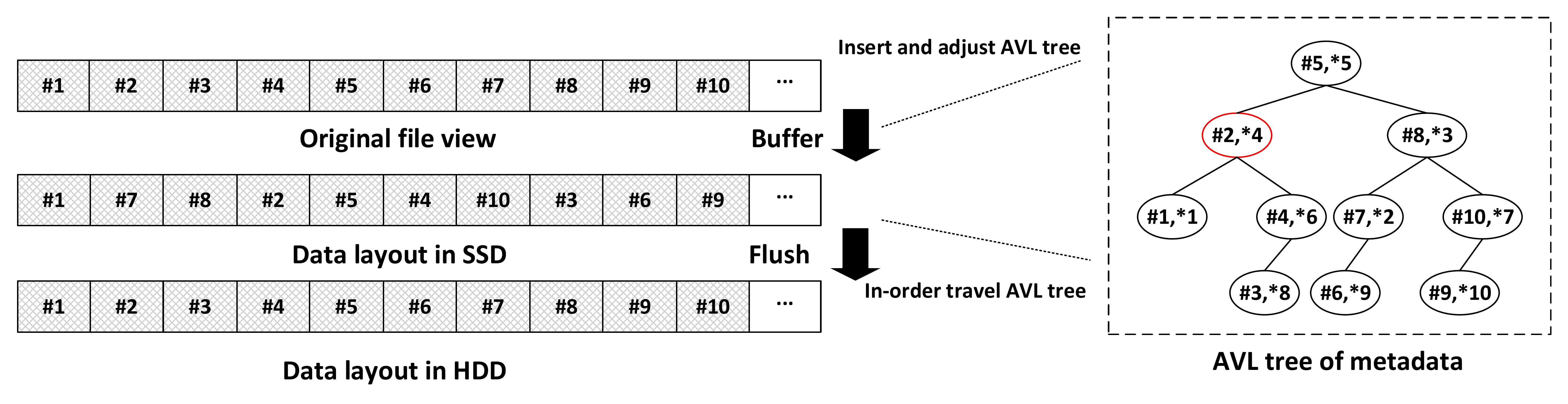}
  \vspace{-0.4cm}
  \caption{Metadata management with the AVL tree}
  \label{avl}
  \vspace{-0.1cm}
\end{figure*}

In order to reduce the performance loss caused by the write amplification~\cite{tang2015ripq} of SSD, SSDUP+ is designed to write the data to SSD in a log-structured way, i.e., append the data to the end of the cached files. However, in the log-structured mode, the original sequence of
the requests are disrupted. As shown in Figure~\ref{avl},
the original data \{\#1,\#2,\#3,...,\#10\} is requested in sequence \{\#1,\#7,\#8,...,\#9\}, which is then the order in which the data is written to SSD.

To recover the original data order and maintain the sequence, we need a mechanism
to manage the metadata of the cached data. Normally, a hash table is a desired choice
because of its $O(1)$ time complexity for queries. In SSDUP+, as we aim to quickly write the
disrupted data back into the HDD, we have to re-sort the cached data.
A typical sorting algorithm like quicksort takes the time of $O(nlog n)$. In this work, we
choose to use the AVL tree to manage the cached data instead of a hash table. The
AVL tree is a self-balancing binary search tree, which takes the time of $O(log n)$ for
the basic operations. When SSDUP+ writes the data to
the SSD, it records the original metadata (including the original offset
and size) and new metadata (including new offset and size). Both original and new metadata of the same data are stored in one leaf node.
Each value requires 8 bytes, which adds up to 24 bytes for one node.
The AVL tree requires about 3MB storage in our experiments (the file size is 40GB and request size is 256KB).
The nodes are sorted based on the original data offset. As shown in Figure~\ref{avl},
in node (\#2,*4), \#2 represents the original offset while *4 represents the new offset
in SSD. Each AVL tree stores the metadata of one file. This way, the data sequence can be
maintained while buffering, which saves an unnecessary sorting phase.

One significant advantage of using AVL trees in our design is that when the data needs
to be flushed to the disk in the sequential order, SSDUP+ only needs to conduct an ordered
traversal of the AVL tree. Note that when traversing the AVL tree, we can ensure the
data is in its original sequence. However, as the way in which the data is written changes the
original data layout, it is the random read to retrieve the data from the SSD. Since
the SSD has the nearly zero seek delay, the random read from SSD does not hurt the performance. Additionally, writing to the SSD sequentially can avoid the write amplification when the SSD is heavily occupied.

\section{IMPLEMENTATION}
In this work, SSDUP+ is implemented and integrated into the trove module in OrangeFS-2.9.3 which is mainly responsible for data storage on the server side. SSDUP+ consists of four modules: random access detector, data redirector, data management and metadata management.
\par
The random access detector is used to analyze the access pattern of the workload and calculate the randomness. The current method of detecting the data access pattern from multiple processes is mainly carried out by calling the collective MPI-IO operations in an application. This \textit{client-side} method can only analyze the data access pattern for a single application, rather than the mixed load that the storage server faces. Our design records the metadata information(such as logical offset, request size, file handle, etc.) of each request that is submitted the server\textit{pvfs-server}, and calculates the randomness of the request streams based on the metadata. The data redirector is used to indicate the write path for each request. It dynamically generates the threshold based on the historical records over a period of time to determine the directions of the requests. The advantage of this scheme is to make full use of the workload information and dynamically adjust the threshold to achieve the better SSD utilization. To implement this module, we rewrote the related function call stack of OrangeFS and made the write logic as simple as possible. The data management mainly adopts the two-stage pipeline mechanism to realize data storage and migration. This module is completely independent of design and implementation and provides a separate interface. The metadata management uses the AVL tree for matedata storage and access. This module is tightly coupled to the random access detector. The metadata recorded by the random access detector is stored in the metadata management and accessed when the data is refreshed.
In addition, SSDUP+ is transparent to users. It is fully compatible with all user operations in OrangeFS and there is no need to modify any user API when using it. Users only need to use the configuration file to set the parameter(e.g. SSD direction, SSD capacity, etc.). Therefore, it is easy to deploy SSDUP+ to large-scale high-performance computing clusters. 

%\section{Experimental Results and Analyses}
\section{EVALUATION}
We have conducted extensive experiments to validate the design of SSDUP+ and to evaluate
its performance. We present the results and analyses in this section.

\subsection{Experimental Setup}

%In this work, SSDUP is implemented and integrated into the trove module in OrangeFS-2.9.3. SSDUP obtains the parameter settings (e.g. thresholds, SSD directory, SSD capacity, etc.) from the configuration file. When the requests arrive at the trove module, SSDUP intercepts them and performs the optimization described in Section 2.

The experiments were conducted on a cluster of 10 nodes, in which 8
nodes are compute nodes, and 2 nodes are I/O nodes.
Each compute node is equipped with 16 Intel Xeon E5-2670 CPU processors, 64GB RAM,
and a 300GB SATA hard drive. Each I/O node is equipped with 16 Intel Xeon E5-2670
CPU processors, 8GB RAM, a 300GB SATA hard drive (Toshiba model MBF2300RC) and a
240GB SSD (INTEL® SSD DC S3520 SERIES). Each node runs CentOS with the Linux kernel version 2.6.32. The default I/O scheduler for HDD is CFQ with a queue size of 128 while the default I/O scheduler for SSD is NOOP~\cite{noop}. All nodes are connected via a Gigabit Ethernet. MPICH-3.0.2 release~\cite{mpich}, compiled with ROMIO, is installed on the compute nodes. 
\par
We focus on comparing SSDUP+ with two contemporary file systems: the original OrangeFS-2.9.3 and OrangeFS with Burst Buffer being integrated. In addition, we also compare SSDUP+ with the previous version of this work, SSDUP~\cite{xuanhua2017ssdup}. OrangeFS adopts a client/server model. It has been widely used as both an experimental platform and a production platform in HPC areas. Files are striped across multiple I/O nodes to enable parallel I/O with high aggregate~\cite{orangefs}. In our experiments, we deployed OrangeFS on all I/O nodes to manage the server-side SSDs as a generic remote-share Burst Buffer which we called OrangeFS-BB. In addition, to assess the potential of SSDUP+ for HPC applications, three typical benchmarks in HPC I/O area are evaluated: IOR, HPIO and MPI-TILE-IO.

%The default stripe size was set to be 64KB.

\subsection{IOR Benchmarking and Analysis}
 
We used IOR-2.10.3, a parallel file system benchmark developed at Lawrence
Livermore National Laboratory, as one of the evaluation benchmarks. IOR provides several
APIs: HDF5, MPI-IO, and POSIX. The MPI-IO API was used in our experiments.
IOR can also run with different access patterns. We conducted three sets of
test using three different access patterns to simulate different execution stages. The first, second and third test set use the access patterns of \textit{segmented-contiguous}, \textit{strided} and \textit{segmented-random}, respectively.

The size of each I/O request is 256 KB and the first two sets of test
write the data to a shared 16GB file. The final set of tests write the data to a shared 8GB file because a completely random pattern is relatively rare. 
%The default number of processes is 32 unless otherwise specified.

\subsubsection{Performance with different numbers of processes.} In the first
set of evaluation, we ran the IOR instances(segmented-contiguous, strided, and segmented-random patterns) with 8, 16, 32, 64, 128, 256 and 512 processes, aiming to investigate the performance and the SSD usage of SSDUP+, the original OrangeFS, OrangeFS-BB, and SSDUP. In this set of experiments, the capacity of SSD is set to be large enough to hold all the data.

\begin{figure*}[!t]
  \begin{minipage}[t]{0.32\linewidth}
    \centering
    \includegraphics[width=\textwidth, height=4.0cm]{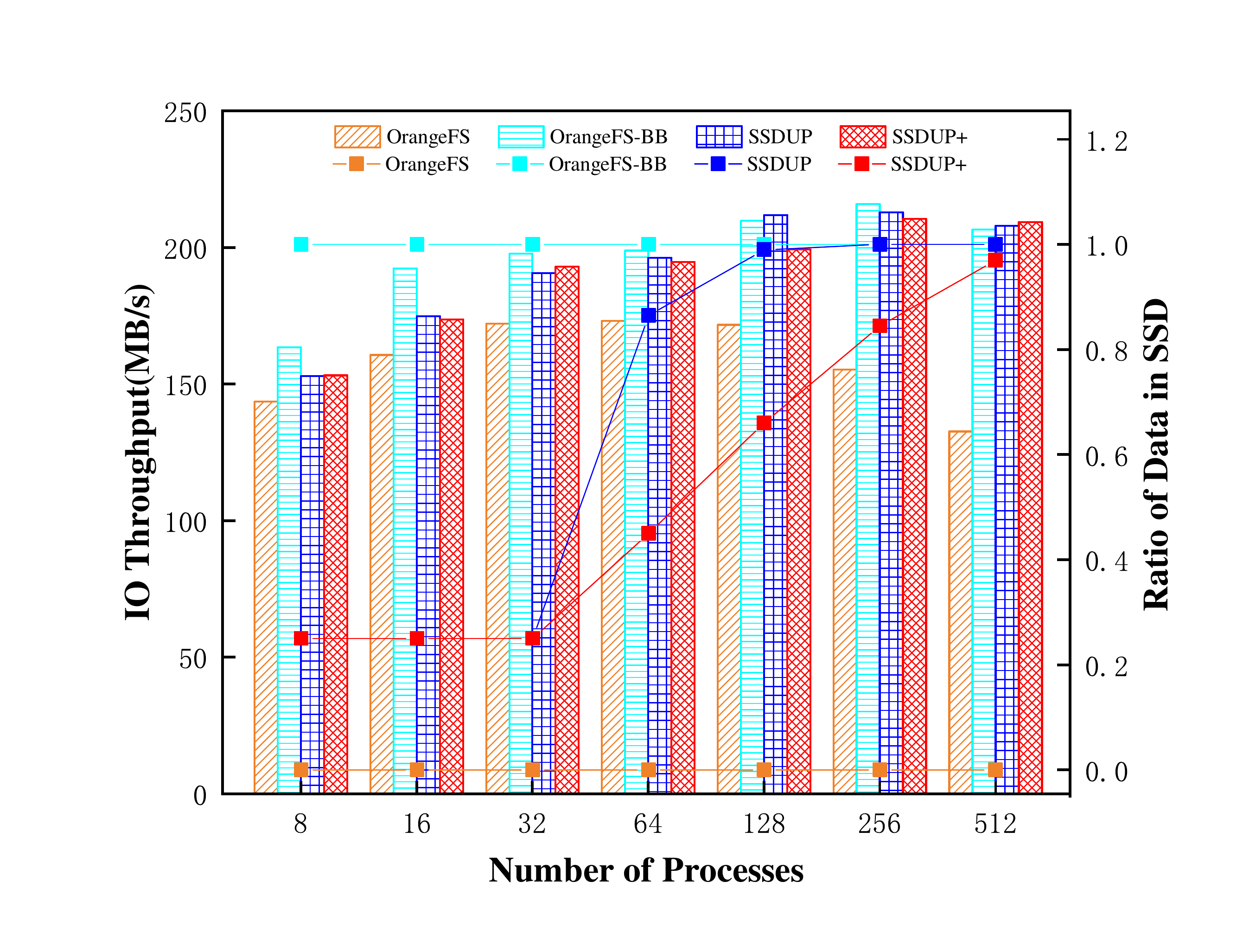}
    \caption{I/O throughput with different numbers of processes for the IOR benchmark, with the original OrangeFS , OrangeFS-BB, SSDUP and SSDUP+ \label{figure-ior-process}}
  \end{minipage}
  \hspace{0.3cm}
  \begin{minipage}[t]{0.31\linewidth}
    \centering
    \includegraphics[width=\textwidth, height=4.0cm]{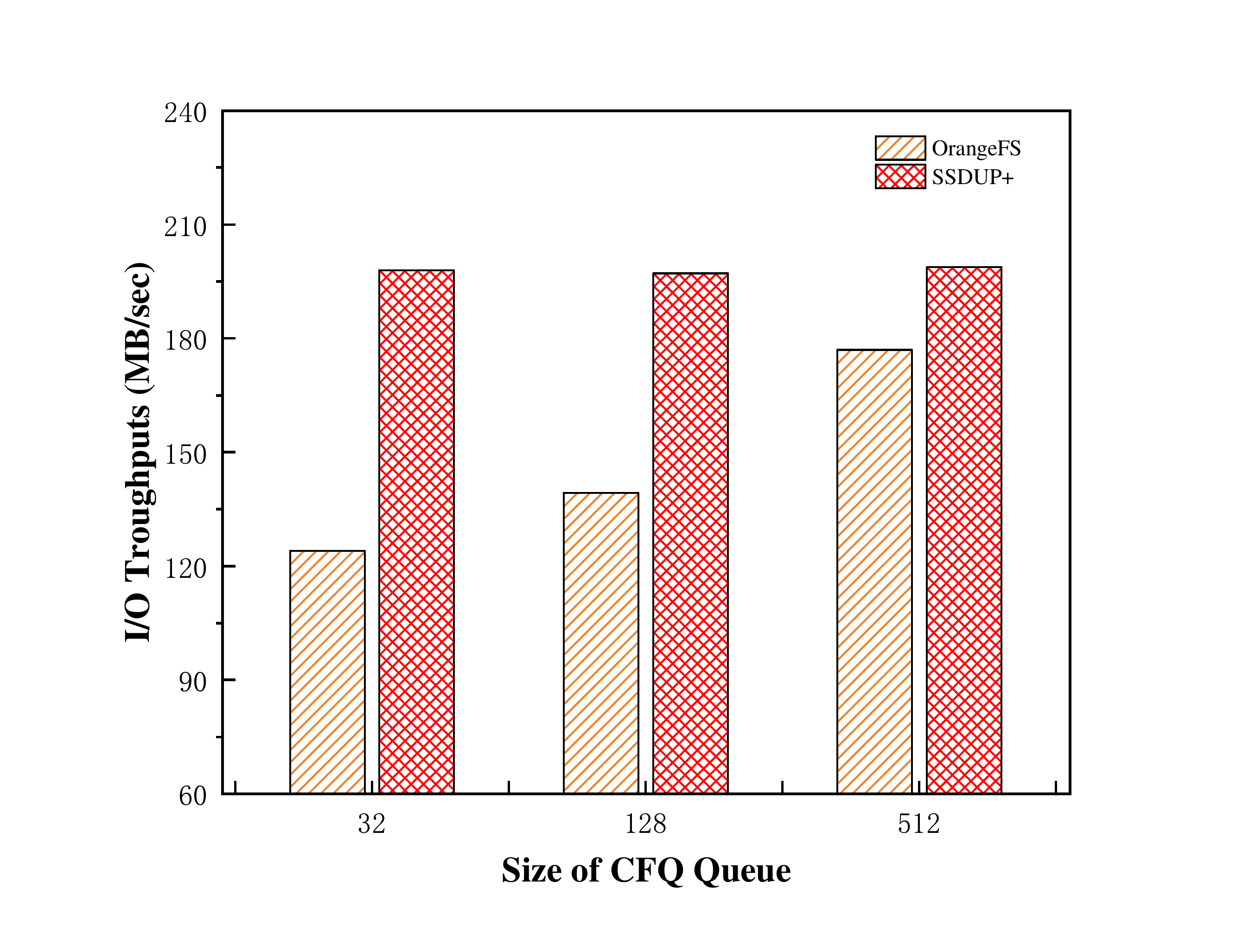}
    \caption{I/O throughput with different sizes of CFQ queue for the IOR benchmark, for the original OrangeFS and SSDUP+ \label{figure-cfq}}
  \end{minipage}
  \hspace{0.3cm}
  \begin{minipage}[t]{0.32\linewidth}
    \centering
    \includegraphics[width=\textwidth, height=4.0cm]{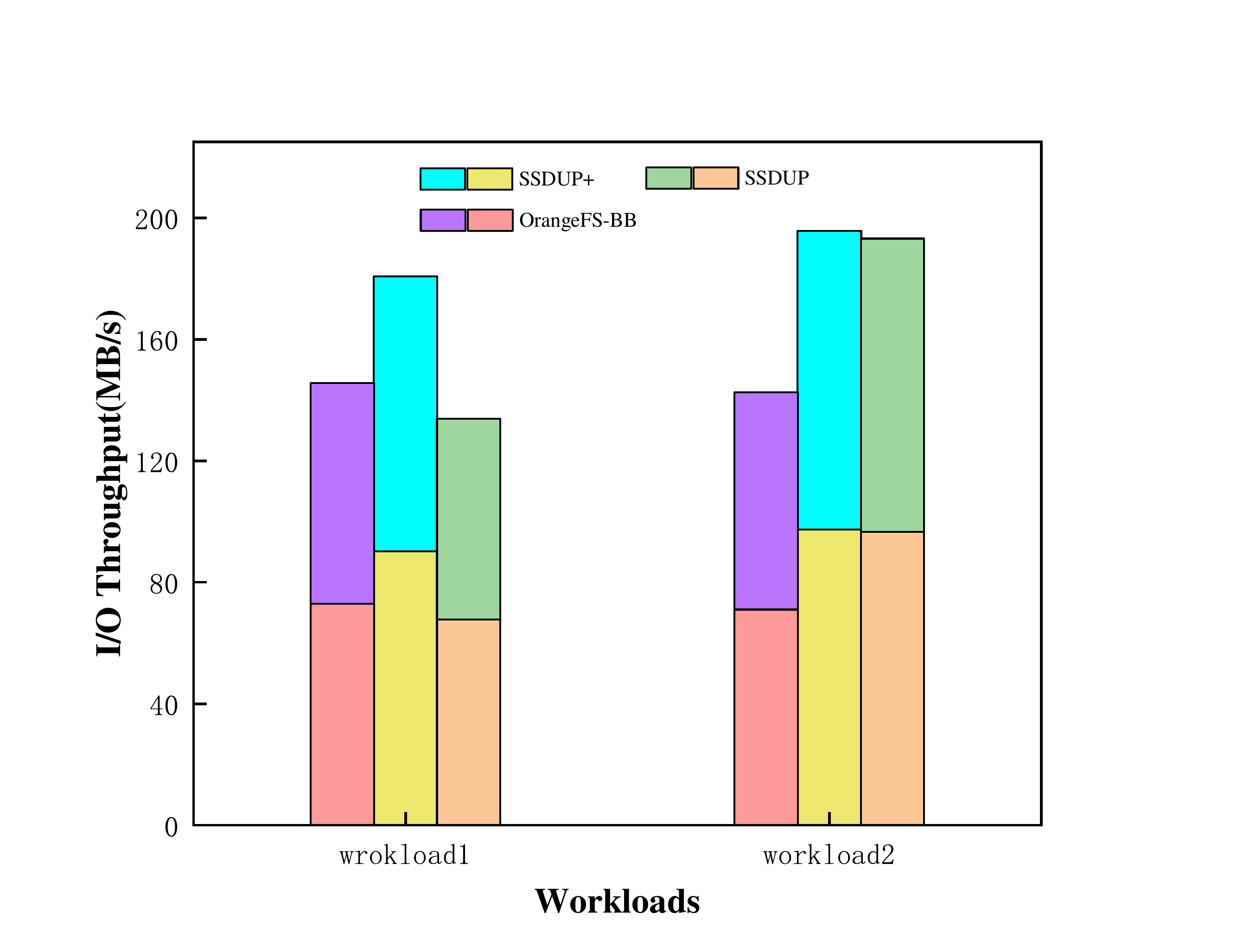}
    \caption{Performance comparison between OrangeFS-BB, SSDUP and SSDUP+ with limited SSDs capacity \label{flush-strategy}}
  \end{minipage}
  \hspace{0.3cm}
  \begin{minipage}[t]{0.32\linewidth}
    \centering
    \includegraphics[width=\textwidth, height=4.0cm]{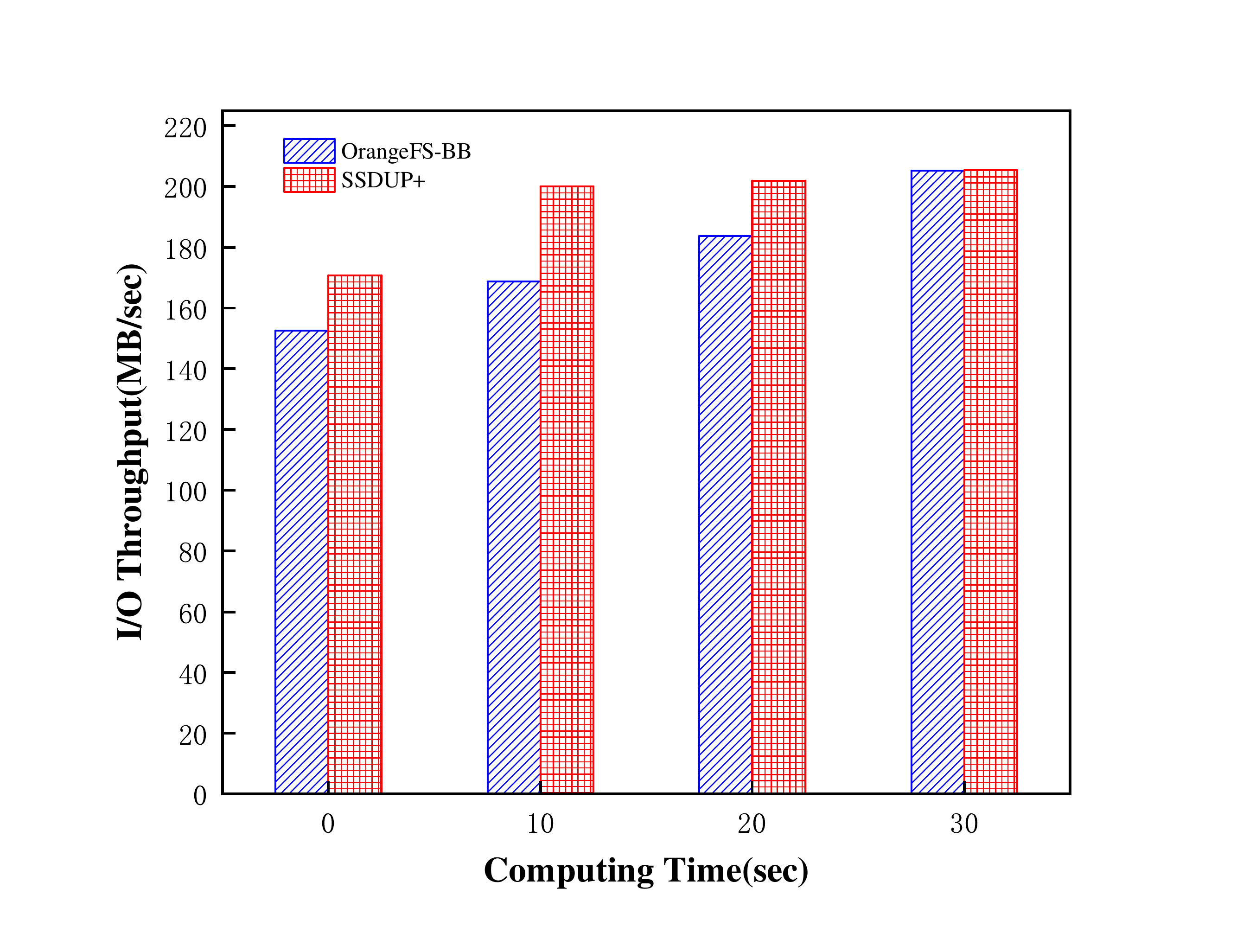}
    \caption{Performance comparison between SSDUP+ and OrangeFS-BB with limited SSDs capacity and different computing time \label{figure-diffcomtime}}
  \end{minipage}
\end{figure*}

Figure~\ref{figure-ior-process} shows the impact of the number of processes on the performance and the SSD usage of SSDUP+. As the number of processes increases from 8 to 32, the performance of the original OrangeFS increases slightly because of the positive effect of parallel I/O. However, as the number of processes increase further (from 64 to 512), the performance of the original OrangeFS starts to decline, which is due to the increased I/O competition caused by too many processes. The performance of OrangeFS-BB shows a different trend. Its performance improves greatly from 8 to 16 processes, then improves slightly from 32 to 512 processes, and eventually maintains at the high performance. This is because OrangeFS-BB uses SSD to cache all data.

The performance of SSDUP+ improves gradually as the number of processes increases from 8 to 32. It reaches the same performance as the original OrangeFS with 32 processes. Moreover, SSDUP+ only buffers 25\% of the data in SSD, which is 75\% less than OrangeFS-BB. This is because SSDUP+ uses the \textit{random factor} to filter the workloads and only buffer random requests in SSD. When the number of processes is small, only the requests with the segmented-random pattern are redirected to SSDs, which account for 25\% of the total data size.

From 64 to 512 processes, the performance of SSDUP+ always maintains at almost the same level as OrangeFS-BB(only  2.15\%, 4.99\%, 2.53\% lower than OrangeFS-BB, respectively). In contrast, SSDUP+ only buffers 40\%, 66\%, 84.5\%, and 97\% of the data, which is less than OrangeFS-BB by 55\%, 34\%, 15.5\%, and 3\%, respectively. This is because when the number of processes increases, more requests with the segmented-contiguous and the strided pattern are identified as random requests due to the I/O interference between the different processes~\cite{zhang2010interferenceremoval}, and therefore are redirected to SSD.

In addition, although SSDUP is almost comparable to SSDUP+ in performance when the number of processes increases to and beyond 64 processes, SSDUP uses much more SSD space than SSDUP+ to cache the data. As shown in figure~\ref{figure-ior-process}, SSDUP+ uses 41.5\%, 33\%, 15.5\% and 3\% less SSD space than SSDUP. This is because SSDUP uses a static empirical threshold to determine whether the requests are random, but SSDUP+ uses an adaptive algorithm to identify the random requests.

\subsubsection{Performance with different CFQ queue sizes.} We also measured the impact of the CFQ queue size on the SSDUP+ performance. We conducted three groups of experiments
with the size of the CFQ queue being 32, 128, and 512. The IOR instances was run with 32 processes.
SSDUP+ achieved 59.7\%, 41.5\%, and 12.3\% performance improvement, respectively,
as shown in Figure~\ref{figure-cfq}.
Note that the default size of the CFQ queue is 128. When the size was changed
to 32, CFQ became more sensitive to the concurrent accesses with interferences,
which resulted in the decrease of the system I/O bandwidth to 124 MB/s.
We also adjusted the length of the request stream to increase the percentage
of random factor, which caused more data to be identified as random accesses
and be redirected to SSD.
The ratio of data directed to SSD became 92\%, which is greater than the ratio in the case where the queue size is
128.
When the CFQ queue size became 512, the aggregate throughput of the original system
increased to 179 MB/s. With a larger queue size, the CFQ scheduler has more opportunities
to merge adjacent requests and achieves better locality. Just because of this reason, the throughput is less sensitive to random accesses ~\cite{zhang2012itransformer}. In these cases, SSDUP+ only
achieved 12.3\% improvement and a small portion of segmented-random data were redirected to SSD.
%the specific ratio is 17.4\%.

\subsubsection{Performance with Limited SSD Capacity}
In order to verify the performance improvement of different pipeline strategies in multi-load scenarios, we tested the performance of different loads under different flushing strategies with the SSD of a small capacity. The reason why we chose a small SSD is because in the HPC area, the amount of data generated in a computing phase is very likely to exceed the Burst Buffer size. Therefore we would like to test the performance of SSDUP+ in the scenarios where the SSD capacity is constrained. 

First, we set the SSD capacity as 8GB. Note that in OrangeFS-BB, the 8GB is used as an entire space, while in SSDUP+, SSD is divided into two 4GB regions. Then we run the following workloads in the tests: $workload_1$ includes two IOR instances and its access pattern is segmented-contiguous and random; $workload_2$ also contains two IOR instances but its access pattern is segmented-random. Each IOR instance writes 8GB of data, and the request size is 256KB.
As shown in Figure~\ref{flush-strategy}, OrangeFS-BB achieves the I/O bandwidth of 73.04MB/s and 72.71MB/s for these two IOR instances in $workload_1$. However, SSDUP+ achieves the bandwidth of 90.21MB/s and 90.49MB/s for two IOR instances, which accounts for 23.98\% of improvement. This is because SSDUP+ uses the two-stage pipeline buffering strategy. When one region is full, the random requests are served by the other region. In this test, SSDUP+ performed three flush operations. The total delay of the first two flush operations was 17 and 19 seconds, respectively. However, due to the SSD being full, OrangeFS-BB can only write the data to HDD while SSD is flushing the data to HDD, which leads to intense I/O contention in HDD between the two streams of data writing. This is one of the reasons why the performance of SSDUP+ is better than OrangeFS-BB. The percentages of $RF$ of $workload_1$ is about 70\%, which suggests that a portion of requests in the workloads are still handled by HDD. It is not always good to perform the flush operation immediately when a region of SSD is full. The flush operation should be performed when the load in HDD is low, so that the I/O contention and consequently the performance loss can be reduced. Since SSDUP+ can accurately determine the write path of each request stream, it can accurately identify the current load in HDD. Besides, the I/O bandwidth achieved by SSDUP+ for two IOR instances of $workload_2$ is 97.32MB/s and 98.38MB/s, which is 8.3\% higher than that of $workload_1$. This is because the requests of $workload_2$ are all random, and there are almost no requests written directly to HDD. When a region of SSD is full, the strategy of flushing immediately to HDD is used, and no delay is required. The performance of $workload_2$ are 71.16MB/s and 71.54MB/s respectively, which are almost the same as $workload_1$.
\par
In addition, we also tested SSDUP, which does not have the traffic-aware flushing scheme. As shown in Figure~\ref{flush-strategy}, SSDUP achieved the I/O bandwidth of 67.85MB/s and 66.15MB/s for the two IOR instances of $workload_1$, which is 34.8\% lower than the bandwidth achieved by SSDUP+. This is because that SSDUP starts flushing immediately once a region of SSD is full without taking into account the current load in HDD, which may lead to intense I/O contention. OrangeFS-BB achieves slightly better performance for $workload_1$ than SSDUP, because OrangeFS-BB uses SSD to handle all data while much less data are handled by SSD in SSDUP (hence a SSD of a much small capacity is needed by SSDUP). 

However, the performance of $workload_2$ is the same in SSDUP+ as in SSDUP, because the access pattern of $workload_2$ is very random and consequently almost all data are written to SSD. The workload written to HDD directly is almost zero. Therefore the flush operation will not experience interference, which is the similar effect as when $workload_2$ is processed by SSDUP+.

%\vspace{-3em}
%\subsubsection{advantage of AVL tree}
\subsubsection{Performance with different computing times}

We have also investigated the impact of different computing times on the system performance.
When the SSD capacity is less than the amount of data being written, which is very common in practical applications, the I/O operations of the applications and the flushing operating of SSD will collide, resulting in performance degradation. Traditional Burst Buffer attempts to solve this problem by overlapping the computing stage and the flushing stage. However, the computing time is difficult to predict. Therefore, we would like to evaluate the impact of different computing times between two consecutive I/O phases on the overall performance of OrangeFS-BB and SSDUP+. We ran two identical IOR instances sequentially and adjusted the interval times between these two instances from 0s to 30s. Each IOR instance was executed with the segmented-random access pattern and produced a 8GB shared file. Moreover, the capacity of one region of SSD was set to 2GB in an I/O node managed by SSDUP+, and the capacity of SSD was set to 4GB in an I/O node managed by OrangeFS-BB. The total size of SSD accounts for 50\% of the size of data to be written.
\par
As shown in Figure~\ref{figure-diffcomtime}, the I/O throughput of OrangeFS-BB is improved gradually when the computing time increases. This is because after the first IOR instance fills the SSD buffer, OrangeFS-BB starts the flushing phase, during which the SSD buffer cannot process the new requests. When the flush phase is over, the SSD buffer can continue to process the second IOR instance. So when the interval between the two IOR instances is very short, the second IOR instance conflicts with the flush phase, causing a sharp drop in performance. 
SSDUP+ outperforms OrangeFS-BB by 11.91\%, 10.65\% and 9.92\% with different computing times. In addition, when the computing time is 0 second, the performance of SSDUP+ is only 20\% lower than the peak performance, while that of OrangeFS-BB is 34\% lower. Especially, the performance of SSDUP+ with a computing time of 10 seconds reaches the peak performance of OrangeFS-BB when the computing time is 30 seconds. This indicates that SSDUP+ can tolerate the increase in computing time much better than the traditional Burst Buffer. 

%\subsubsection{blktrace}

\subsection{HPIO Benchmarking and its Analysis}
\begin{figure*}[!t]
  \begin{minipage}[t]{0.32\linewidth}
    \centering
    \includegraphics[width=\textwidth, height=4.0cm]{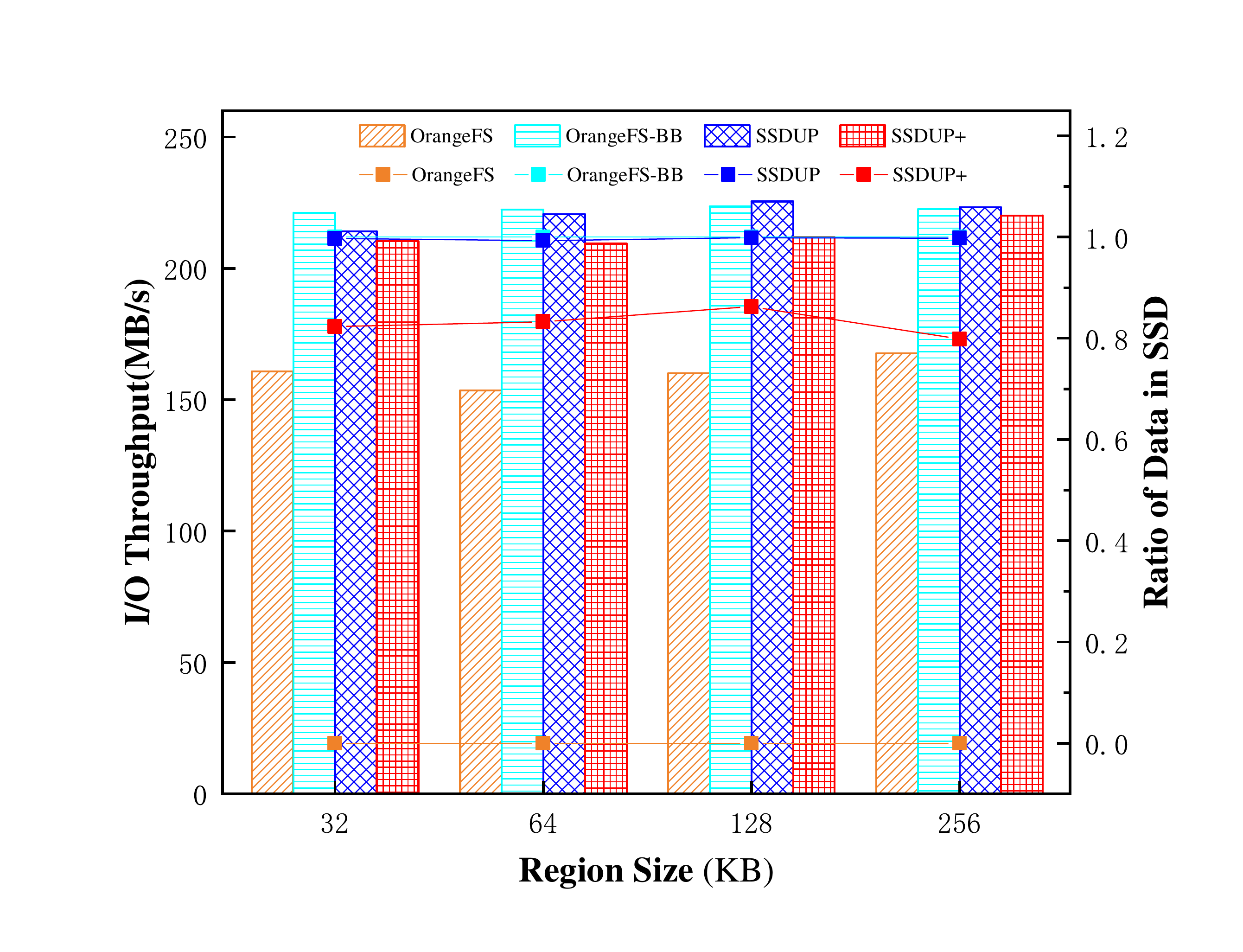}
    \caption{I/O throughput with different region size for the HPIO benchmark, with the original OrangeFS, OrangeFS-BB, SSDUP and SSDUP+ \label{figure-hpio}}
  \end{minipage}
  \hspace{0.3cm}
  \begin{minipage}[t]{0.31\linewidth}
    \centering
    \includegraphics[width=\textwidth, height=4.0cm]{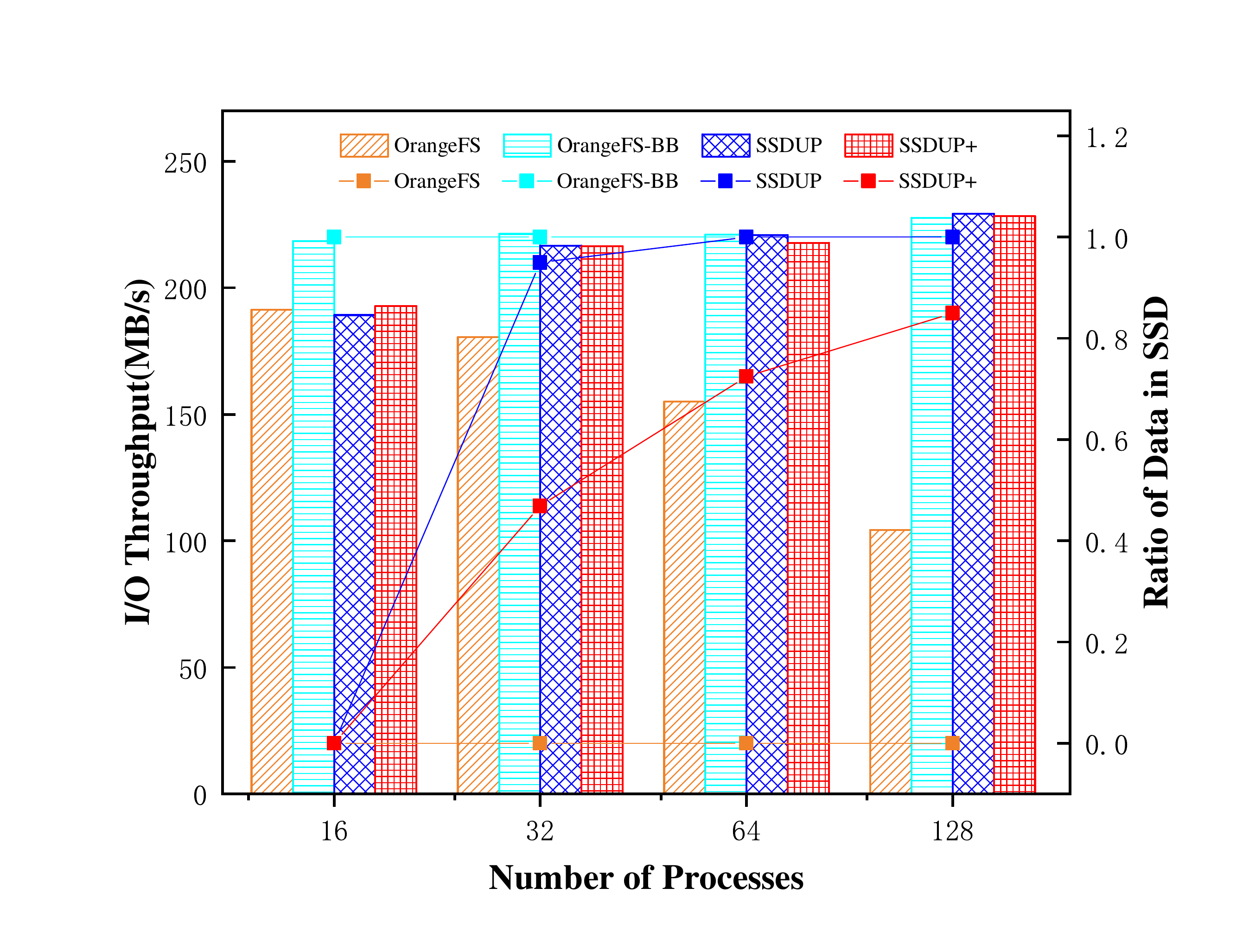}
    \caption{Comparing the original OrangeFS, OrangeFS-BB, SSDUP and SSDUP+ in I/O throughput of the MPI-Tile-IO Benchmark with different number of processes \label{figure-mpi-tile-io}}
  \end{minipage}
\end{figure*}

We also used the HPIO benchmark~\cite{ching2006hpio} to evaluate and compare SSDUP+, the original OrangeFS, OrangeFS-BB and SSDUP in terms of both I/O throughput and the SSD usage. HPIO is a benchmark developed at the Northwestern
University and has been widely used to evaluate the performance of non-contiguous I/O accesses.
Similar to IOR, HPIO can be run in different access patterns too. Four parameters
were used in this test: region size, region count, region spacing, and non-contiguous test array. A region is a piece of contiguous data in a file that will be accessed by a process. The region count is the number of regions accessed by a process. Region spacing is the distance between two adjacent regions. Non-contiguous test array indicates whether the file access is continuous or random. 

We set the region size from 32KB to 256KB and the number of processes is set to 32. The region count varied from region size in order to keep the file size to be around 8GB. In addition, we set the region space as 0. In order to fully evaluate the potential of SSDUP+, we run two HPIO instances concurrently to simulate complex workloads, one of which is of the continuous pattern with the non-contiguous test array set to 1000 (c-c). The other is of the random pattern, and the non-contiguous test array was set to 0010(c-nc). 

As shown in figure~\ref{figure-hpio}, the performance of OrangeFS-BB and SSDUP is roughly the same because both buffered 100\% of the data in SSD. The performance of SSDUP+ is slightly lower than that of SSDUP, but not by more than 6\%(actually 2.84\%, 5\%, 5.2\%, 1.4\%, respectively). This is because SSDUP+ cached less data in SSDs than SSDUP, and the saved SSDs space is 17.39\%, 16.07\%, 13.6\%, 19.86\%. This means that in this scenario, SSDUP+ sacrifices less than 6\% of the performance compared with OrangeFS-BB and SSDUP, but saves an average of more than 15\% of the SSD space.

\subsection{MPI-Tile-IO Benchmarking and Analysis}

MPI-Tile-IO benchmark~\cite{mpi-tile-io} is a member of the Parallel I/O Benchmarking Consortium, which is a test application that implements the tile access to a two-dimensional dense dataset.
Each process accesses a tile, the size of which is based on the number of elements in a tile and the size of one element.
In this test, we run two MPI-Tile-IO instances concurrently with 16, 32, 64 and 128 processes. The first instance was set as a one-dimensional dense dataset whose \textit{x} direction was set to 1 and the \textit{y} direction set to the number of processes. In the second instance, as a normal two-dimensional dataset, the \textit{x} direction is set as the square root of the number of processes, while the product of the \textit{y} direction and \textit{x} direction is kept to the number of processes.
In addition, the size of each element is set to 4KB and each instance generated 16GB of data.

As shown in Figure~\ref{figure-mpi-tile-io}, the I/O throughput achieved by OrangeFS drops as the number of processes increases. This is because the I/O contention between different MPI-TILE-IO instances increases as the number of processes increases, resulting in a much greater randomness for mixed loads.
OrangeFS-BB maintains the peak performance by writing all data directly to SSD. SSDUP+ and SSDUP have the same performance as OrangeFS when there are 16 processes. This is because there is no request identified as the random request, and therefore the SSD space used is 0. When the number of processes is 32, the performance of SSDUP+ and SSDUP is almost the same as OrangeFS-BB, but only 46.87\% of the data is written to SSD in SSDUP+, while 95\% of the data is written to SSD in SSDUP. Nearly 50\% of the SSD space is saved in SSDUP+.
As the number of processes increases further, SSDUP identifies all data as random requests and writes them to SSD, just as in OrangeFS-BB. However, SSDUP+ saves 27.5\% and 15\% of SSD space, respectively, comparing with SSDUP and OrangeFS-BB, and always maintains as high performance as OrangeFS-BB.

\subsection{Overhead Analysis}

The overhead of SSDUP+ primarily comes from two aspects: the cost of grouping
and sorting the requests in a request stream (called \textit{grouping cost}), and
the cost of keeping the AVL tree balanced and in order and travelling the AVL tree
(called \textit{AVL cost}). We use the IOR benchmark to study and analyze
these overheads. The total size of the accessed data is 2GB, and the request size
varies from 32KB to 512KB. The SSD capacity was set to be 2GB. The IOR benchmark
was executed with the segmented-random pattern, and all requests were sent to the SSD.

The system overheads was measured to be 0.13\% (in the case of 512KB requests)
and 0.79\% (in the case of 32KB requests) of the total execution time, which can be ignored compared to the performance gain. As shown in
Table 2, these two types of overhead increase as the request size becomes smaller.
This is because when the request size becomes smaller, the number
of I/O requests increases. The overheads
in the cases of 128KB and 64KB requests are
 close, because the request is striped
across two data servers when the request is larger than the default stripe size.

%group cost is larger than AVL cost

These experimental results show that SSDUP+ improves the write performance
when using 40\% of the total SSD space (the ratio of random
accesses is 20\%) for the IOR benchmark, and 50\% of the total SSD space
(the ratio of random accesses is around 20\% to 100\%) for the HPIO benchmark, and 90\%
of total SSD space (the ratio of random accesses is in the range of 80\% to 95\%) for the MPI-Tile-IO benchmark.
The overhead is negligible, less than 1\% of the total execution time.
The average ratio of random accesses in HPC applications has been reported to be
around 50\%~\cite{zhang2013ibridge}, which is consistent with the ratio setting
in our experiments. Compared with burst buffer, SSDUP+ can save up to 50\% SSD space. In addition, our approach does not introduce more writes to SSD. Instead, SSDUP+ only buffers random I/O requests and reduces the number of writes to SSD. As the result, SSDUP+ can extend the lifetime of SSD compared with the design of conventional burst buffer. Since IOR, HPIO, and MPI-Tile-IO are typical benchmarks that represent the common access patterns (segmented-contiguous, segmented-random, strided, noncontiguous and nested-stride) of scientific applications, we believe that SSDUP+ can achieve similar performance gains for realistic workloads.

\begin{table}
  \renewcommand{\arraystretch}{1.3}
  \caption{System Overhead}
  \label{table_overhead}
  \centering
  \begin{tabular}{|c|c|c|c|}
    \hline
    \bfseries Request Size & \bfseries Total Time & \bfseries Group Cost & \bfseries AVL Cost \\
    \hline
    32 KB & 15.5 s & 29.1 ms & 93.4 ms \\
    \hline
    64 KB & 12.5 s & 12.5 ms & 44.2 ms \\
    \hline
    128 KB & 11.8 s & 15.9 ms & 41.8 ms \\
    \hline
    256 KB & 11.54 s & 9 ms & 21.2 ms \\
    \hline
    512 KB & 11.9 s & 6.1 ms & 9.5 ms \\
    \hline
  \end{tabular}
\end{table}

\section{Related Work}

Many research studies have been conducted to improve the performance of the I/O system for HPC
applications~\cite{thakur1999data, yin2013pattern, zhang2012opportunistic}.
In this section, we discuss the existing work in four areas: i) addressing the random access
problem caused by concurrent access to the hard disk, ii) identifying the critical data and access patterns, iii) the storage systems with SSD, and iv) the extension from the previous version of this work, SSDUP.

\subsection{The Problem of Concurrent Access}
Wang et. al. introduced an IBCS method~\cite{wang2014iteration}, based on two-phase I/O, to reorganize the data transferring order in the shuffle stage to keep a one-to-one object storage target access pattern~\cite{dickens2009lib} in each iteration, which prevents multiple processes from competing for one disk head. Because only one process accesses the disk at a time, the system throughput is significantly improved due to the reduced cost of disk seeking.
Chen et. al.~\cite{chen2010improving} analyzed the difference between the data layout at the client
side and the data layout in parallel file systems. The difference causes the processes to interfere each other when they
access the hard disk. Zhang et. al.~\cite{zhang2010iorchestrator} reported
that the file striping strategy of a parallel file system may jeopardize the locality of individual programs when multiple programs were served concurrently by a data server.
Zhang et. al.~\cite{zhang2010interferenceremoval} studied a common pattern of HPC applications,
that is, a) the offsets of individual processes are continuous; but b) multiple processes may compete for
a disk head when the processes handle the requests concurrently, which makes the disk head
move back and forth. They propose a data replication method that copies the data of the same process to the same I/O node, and ensures that each I/O node serves as few processes as possible.
All these approaches will operate on the data itself (merge or migrate, etc.), and the system we designed does not need to operate on the data, but only analyzes the metadata to obtain the data access pattern. And the historical access record is used to guide the writing of next requests. So our design is simple and efficient, and the system overhead is extremely low.

\subsection{Recognizing Critical Data and Access Patterns}

S4D-Cache~\cite{he2014s4d} introduces a new technique to identify the data that is critical to performance and redirect these data to SSD. S4D-Cache migrates the data between SSD and
HDD according to the temporal locality. The method is able to keep the data with strong temporal locality in the SSD longer. Byna's prefetching technique~\cite{byna2008parallel} prefetches the data into the memory by detecting a stable local access pattern.
Both S4D-Cache and I/O prefetching analyze the access pattern from the perspective of a single process.
However, multiple processes that access the same server can cause competition and random accesses. More processes are involved, more random the IO accesses can be. It is difficult to calculate
the cost accurately from a single process' point of view. Our proposed method traces
the access pattern in a global view, which
can measure the random access caused by both an application's native behavior and the competition between multiple processes.

\subsection{Storage Systems with Burst Buffer}
Burst buffer~\cite{liu2012role} inserts the I/O forwarding nodes equipped with SSD/DRAM in clients or between the clients and
the storage nodes. By doing so, the data is flushed from the memory to the persistent storage quickly and the execution flow can return to the computation
phase as soon as possible.
A number of Burst Buffer systems focus on using the high performance of DRAM to accelerate bursty writing(eg. checkpointing). For example, CRUISE~\cite{rajachandrasekar20131} and BurstMem~\cite{wang2014burstmem} are using distributed memory file system to speed up burst write, but their drawbacks are obvious. On the one hand, small-capacity memory cannot achieve the purpose of acceleration and the cost of large-capacity memory is unbearable; on the other hand, when the system crash or power failure, the data in memory will be directly lost. However, our SSDUP+ is optimized for systems with Burst Buffer whose mainly physical hardware is SSD. The price-to-capacity ratio of the SSD is much lower than that of the DRAM, and the data in SSD is not easily lost when system crash or power off.

%Ligang 0206 1010am
In addition, some works related to the Burst Buffer use SSD as the main cache device. Hystor~\cite{chen2011hystor} identifies the data which cause long latencies and store them in SSD. However, it needs to maintain an overall data map for entire system, which is very expensive. Moreover, Hystor only considers the I/O access patten, but does not take into account the characteristics of applications in high-performance computing. In addition, Hystor is designed only for stand-alone and embedded kernels, not for parallel environment. 
BurstFS~\cite{wang2016burstfs} carefully analyzes the data characteristics of scientific applications to access the data efficiently. But its goal is to pursue the optimal performance, without considering the efficiency and cost-effectiveness of SSD. 

SSDUP+ addresses the problems in the above designs. First, by tracing I/O from a global perspective, we can detect the random accesses caused by various situations, which makes the I/O accesses more efficient. Second, by designing a pipeline mechanism and the AVL tree, we can use the SSD space in a more efficient way, the result of which is to use less space to buffer more requests and not to rely on the accurate prediction of computation time to overlap the flushing stage.

\subsection{Extension from SSDUP}
\label{extension}
The previous version of this work has been published in ICS2017~\cite{xuanhua2017ssdup}. SSDUP+ extends SSDUP from the following aspects. These non-trivial extensions substantially improve the performance of SSDUP.

\par
First, there are various I/O access patterns, such as segmented-contiguous, segmented-random, strided and the mixed pattern. In SSDUP, we analyze the contiguous, random and strided access pattern in detail, but do not consider the mixed pattern, which is generated by multiple applications which issue the I/O operations simultaneously. The mixed access patter is often seen in the real HPC systems and is much more complicated than a single access pattern. In SSDUP+, we analyze the I/O trace of mixed pattern. The fundamental correlation between I/O performance and randomness of the data being accessed is further revealed. Based on these analyses, an adaptive algorithm is developed to accurately identify random requests based on the I/O characteristics. We conduct the experiments to verify the effectiveness of the new design. The results show that after incorporating the adaptive algorithm in our data-redirection module, SSDUP+ is able to save more SSD space than SSDUP.
\par

Second, the two-stage pipeline (i.e., caching and flushing) mechanism is designed to manage the data in SSD. The flushing stage is the key in the pipeline module. In SSDUP, the flushing stage is performed immediately after a region (half of the SSD space) is full. This strategy of immediate flushing works well with the situation where the requests are highly random   (and therefore little data are written to HDD directly). However, as the I/O access pattern become more complex such as in the mixed pattern, the strategy of immediate flushing does not often achieve good performance improvement. This is because when the data is being flushed to HDD, the random data generated by the applications may also be written to HDD at the same time, which we found may cause intense interference between the two streams of data writing and therefore lead to performance degradation. To address this issue, we propose a traffic-aware flushing strategy in SSDUP+. The pipeline module of SSDUP+ dynamically analyzes the current workload and decides a good timing for performing the flushing operation according to the current workload in HDD. We also conduct the experiments to evaluate the traffic-aware pipeline mechanism. Our results show that with the new design, SSDUP+ outperforms SSDUP significantly in terms of I/O throughput, especially when the SSD capacity is not sufficient comparing the amount of I/O data generated by the applications. 
%In prusuit of maximum performance, only random requests are cached in SSD, which saves SSD space to the greatest extent. At the same time, when the capacity of SSD is insufficient to hold all data, the flexible flushing strategy is adopted to flush data while maintain good performance.

%Our system differs from other SSD-HDD systems in two fundamental aspects. First, by tracing I/O from a global point of view, we can detect the random accesses caused by various situations, which makes the I/O accesses more efficient. Second, by designing a pipeline mechanism and the AVL tree, we can use the SSD space in a more efficient way, the result of which is to use less space to buffer more requests and not to rely on the accurate prediction of computation time to overlap the flushing stage.
%preventing SSD from being full

\section{Conclusion}

In this paper, we propose a traffic-aware SSD burst buffer scheme, called SSDUP+.
We carefully analyze three common access patterns in the HPC environment.
The proposed I/O-traffic detection method is able to identify random/irregular I/O accesses caused by various access patterns.
%proposed a notion of \textit{random factor} to identify random/irregular I/O accesses.
SSDUP+ is designed to buffer the random data into SSD. SSDUP+ is able to identify the random data in the incoming I/O requests based on the proposed metric of \textit{random factor}. Then, an adaptive algorithm is proposed to dynamically adjust the threshold of the random factor. The data with the randomness higher than the threshold will be cached to SSD. Moreover, the requests in SSD are sorted and flushed to HDD using a pipeline mechanism. Given the fact that multiple applications may generate a complex I/O access pattern, a traffic-aware flushing strategy is further designed to reduce the I/O interference between the flushed data and the date being written to HDD directly. In order to maintain the data sequence, an AVL tree data structure is used to manage the buffered data. With all above techniques,
SSDUP+ can achieve the desired I/O performance with a much smaller SSD space.

We have implemented a prototype of SSDUP+ in OrangeFS to validate the design and evaluate its performance benefits. The methodology proposed in SSDUP+ is generic and is applicable in other file systems. Because our scheme is designed to be deployed at the server side, we do not need to consider the I/O patterns of individual applications.

We conduct the experiments with three widely-used benchmarks, IOR, HPIO, and MPI-Tile-IO. The experimental results show that SSDUP+ can save 50\% of SSD space on average, while delivering similar performance as other common burst buffer systems. SSDUP+ can also save 20\% of SSD space on average, comparing with SSDUP. Especially, when the SSD capacity is insufficient to accommodate all data, SSDUP+ can improve the performance (I/O throughput) by 20\% to 30\%, comparing with the common burst buffer scheme and SSDUP. 

In near future, we plan to broaden the adoption of SSDUP+. We plan to port SSDUP+ to the Lustre~\cite{lustre} file system, and make full use of SSDUP+ for HPC storage systems. We also plan to apply the design methodologies in SSDUP+ to new devices such as non-volatile memory and new hybrid storage architectures. 
% Start of "Sample References" section
\begin{acks}

This work is supported by the National Key Research and Development Plan (No. 2017YFC0803700), NSFC (No.61772218, 61433019), and the Outstanding Youth Foundation of Hubei Province (No.2016CFA032). 
\end{acks}
% Bibliography
\bibliographystyle{ACM-Reference-Format}
\bibliography{refers-cr}

\end{document}